\documentclass[10pt,journal,compsoc]{IEEEtran}
\IEEEoverridecommandlockouts
\usepackage[T1]{fontenc}
\usepackage[utf8x]{inputenc}
\usepackage{cite}
\usepackage{amsmath,amssymb,amsfonts}
\usepackage{graphicx}
\usepackage{textcomp}
\usepackage{balance}
\usepackage{xspace}
\usepackage{multirow}
\usepackage{float}
\usepackage{soul}
\usepackage{microtype}
\usepackage{url}
\usepackage{booktabs}
\usepackage[dvipsnames]{xcolor}
\usepackage{listings}
\usepackage{paralist}
\usepackage[frozencache,cachedir=.]{minted}
\usepackage{enumitem}
\usepackage[compatibility=false]{caption}
\usepackage{subcaption}
\usepackage{wrapfig}
\usepackage{algorithm}
\usepackage[noend]{algpseudocode}
\usepackage{lineno}

\def\BibTeX{{\rm B\kern-.05em{\sc i\kern-.025em b}\kern-.08em
		T\kern-.1667em\lower.7ex\hbox{E}\kern-.125emX}}

\newcommand{\deleted}[1]{%
	{}}%

\newcommand{\added}[1]{%
	{#1}}%

\newcommand{\changetxt}[2]{{\color{black}#2}}
\newcommand{\changetxtNoRev}[1]{{\color{black} #1}}

\newcommand{\mydef}[1]{%
	\textbf{#1}}%

\begin{document}
	
	\title{Automated Probe Life-Cycle Management for Monitoring-as-a-Service}
	
	\author{Alessandro~Tundo,
		Marco~Mobilio,
		Oliviero~Riganelli,
		and~Leonardo~Mariani,~\IEEEmembership{Senior~Member,~IEEE}
		\IEEEcompsocitemizethanks{\IEEEcompsocthanksitem The authors are with the Department of Informatics, Systems and Communication (DISCo), University of Milano-Bicocca, Milan,
			Italy.\protect\\
			E-mail: \{alessandro.tundo, marco.mobilio, oliviero.riganelli, leonardo.mariani\}@unimib.it}
	}
	
	\maketitle
	\IEEEpeerreviewmaketitle
	
	\begin{abstract}
		Cloud services must be continuously monitored to guarantee that misbehaviors can be timely revealed, compensated, and fixed. While simple applications can be easily monitored and controlled, monitoring non-trivial cloud systems with dynamic behavior requires the operators to be able to rapidly adapt the set of collected indicators. 
		Although the currently available monitoring frameworks are equipped with a rich set of probes to virtually collect any indicator, they do not provide the automation capabilities required to quickly and easily change \added{(i.e., deploy and undeploy)} the probes used to monitor a target system. Indeed, changing the collected indicators beyond standard platform-level indicators \deleted{is a manual, error-prone and expensive process}\added{can be an error-prone and expensive process, which often requires manual intervention}.
		
		This paper presents a Monitoring-as-a-Service \deleted{architecture}\added{framework} that provides the capability to \emph{automatically} deploy and undeploy arbitrary probes based on a user-provided set of indicators to be collected. The life-cycle of the probes is fully governed by the \deleted{architecture}\added{framework}, including the detection and resolution of the \emph{erroneous states} \changetxt{R1.C2}{at deployment time}. The \deleted{architecture}\added{framework} can be \deleted{exploited}\added{used} jointly with \emph{existing monitoring technologies}, without requiring the adoption of a specific probing technology. 
		
		We experimented our \deleted{architecture}\added{framework} with cloud systems based on containers and virtual machines, obtaining evidence of the efficiency and effectiveness of the proposed solution.
		
	\end{abstract}
	
	
	\newcommand{\alessandro}[1]{\textcolor{Plum}{{\it [Alessandro: #1]}}}
	
	\newcommand{\oliviero}[1]{\textcolor{olive}{{\it [Oliviero: #1]}}}
	
	\newcommand{\marco}[1]{\textcolor{orange}{{\it [Marco: #1]}}}
	
	\newcommand{\leonardo}[1]{\textcolor{blue}{{\it [Leonardo: #1]}}}

	\section{Introduction} \label{sec:introduction}

	Cloud-based solutions emerged as the key paradigm to support the operation of large-scale distributed systems composed of many interconnected services~\cite{CloudCompStats2020,Ferrer:DecentralizedCloud:CSUR:2019}. Indeed, these systems are characterized by highly \emph{dynamic} and \emph{complex} behaviors that include the capability to adapt to changes in the available computational resources, to dynamically update and scale services without interrupting operation, and to be resilient to network problems and software failures.
	
	Due to their size and complexity, every element of a cloud system must be \emph{continuously observed}, to timely react to anomalous behaviors, generating alerts, and activating countermeasures~\cite{romano_novel_2011,cedillo2015monitoringcloud,shatnawi2018chmm,wang2011monitoringdc,kutare2010monalytics}. In fact, cloud-based solutions are systematically enriched with monitoring capabilities, either natively offered by cloud platforms (e.g., Kubernetes~\cite{kubernetes}), or provided by external tools (e.g., Elastic Stack~\cite{elasticsearch_2020_stack} and Prometheus~\cite{prometheus}).
	
	These monitoring solutions are mainly designed to collect a stable set of indicators over time, being \emph{challenged by scenarios that require rapidly modifying the set of collected indicators}.
	\changetxtNoRev{In contrast, there are many well-known causes of sudden changes to the set of collected indicators. The \emph{goals of the operators} change with the technical and business objectives of the organization, consequently causing changes in the set of the indicators that must be collected. The \emph{software usage patterns} that emerge from the field continuously evolve, often determining  the need of adjusting the monitored indicators accordingly. The collected indicators must be adapted to changes in the \emph{workload}, which must be carefully observed to timely reveal any symptom of stress on the services. Moreover, \emph{service updates} normally require putting in place ad-hoc monitoring capabilities that target the updated services to measure their reliability and timely detect misbehaviors. Sometimes, the observation of \emph{failures} generates the need of continuously observing the services that fail often, to prevent new failures and localize the causes of problems; and \emph{dynamically deployed scenarios} (e.g., to timely react to disasters and emergencies) require quickly deploying new functional services and the corresponding monitoring components.}
	Relevantly, all these factors are \emph{dynamic and cannot be entirely anticipated}.
	
\begin{figure*}[ht]
		\centering
		\includegraphics[width=0.75\linewidth]{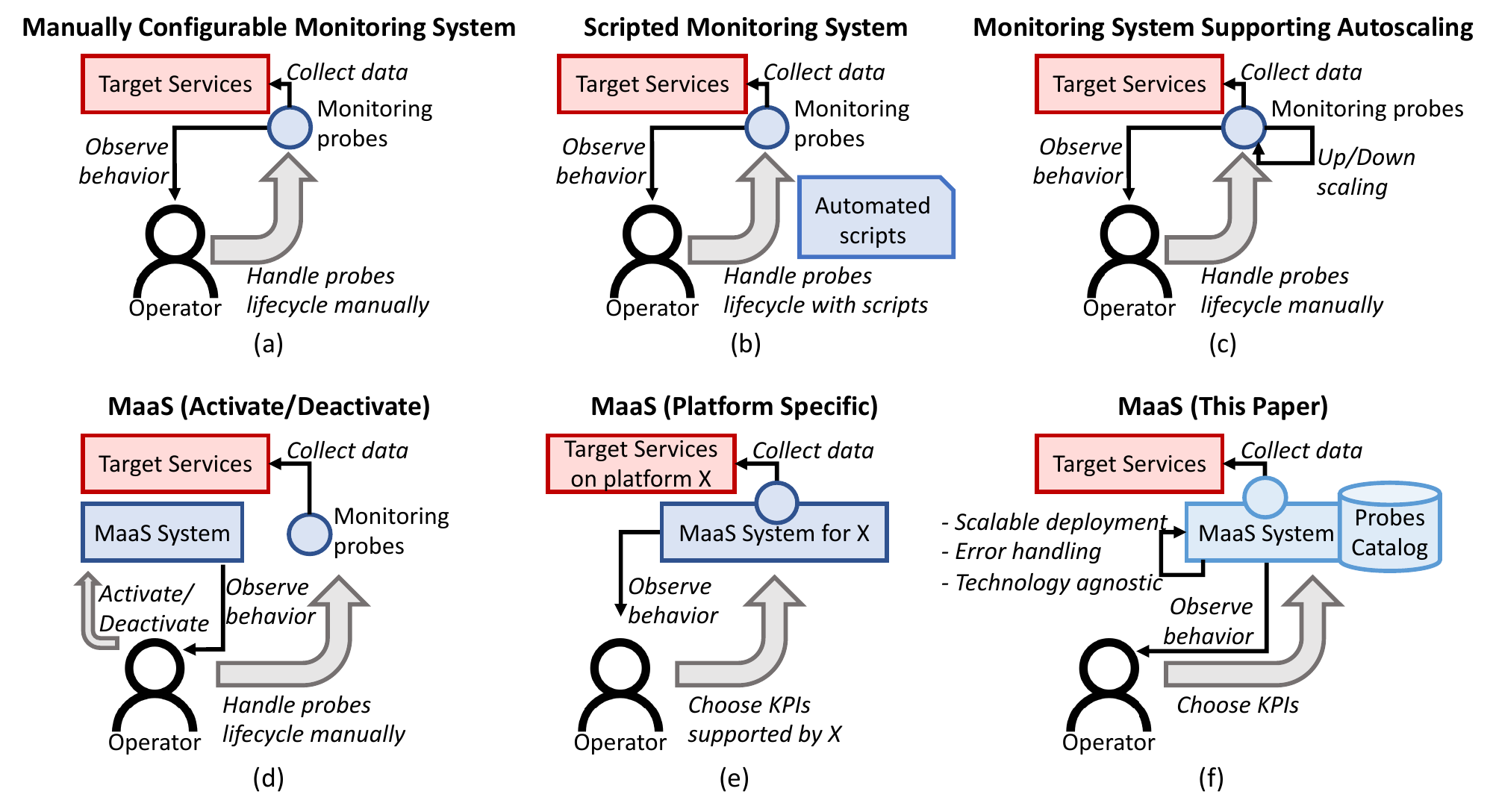}
		\caption{Monitoring automation.}
		\vspace{-1.5em}
		\label{fig:monitoringAutomation}
	\end{figure*}

	Changing the set of collected indicators often requires changing the set of probes running in the field. However, configuring and deploying new probes, as well as undeploying the existing probes, are \emph{non-trivial and time-consuming activities}.
	\changetxt{R1.C3}{For instance, a tech company running many cloud services needs to collect KPIs at different granularity levels, taking into account both business and technical needs~\cite{shatnawi2018chmm}. The needs of managers shall follow business goals and market evolution, while the needs of technicians shall follow QoS goals and software evolution. These needs evolve independently, and simultaneous changes in both business and technology may generate a rapidly increasing number of requests for the operators responsible of configuring the monitoring system.  
		Operators may struggle adapting their monitoring systems at some point, especially when a large number of services has to be monitored.}
	For this reason, research focused on \emph{increasing the level of automation of probe management}. Figure~\ref{fig:monitoringAutomation} shows the increasing levels of automation that have been introduced in monitoring systems. 
	
	%

	Simple \emph{manually configurable} monitoring systems (Figure~\ref{fig:monitoringAutomation} (a)), such as Elastic Stack~\cite{elasticsearch_2020_stack} and Prometheus~\cite{prometheus}, require configuring and deploying probes manually, that is, the life-cycle of every component of the monitoring system must be handled manually by developers. Although useful, these monitoring systems are expensive to use in presence of frequent changes to the set of collected indicators, and badly adapt to dynamic scenarios. 
	
	Some probe deployment tasks could be implemented using \emph{general purpose deployment systems} (Figure~\ref{fig:monitoringAutomation} (b)), such as Ansible~\cite{redhat_2020_ansible} and Puppet~\cite{puppet}. However, these systems are not designed to specifically serve monitoring frameworks, and defining and controlling the deployment strategies would still be entirely on the shoulder of the operator. As discussed next in this paper, general purpose deployment systems can be indeed used as basic building blocks of more sophisticated deployment solutions tailored to address cloud monitoring.

	A simple form of automation present in some systems consists of the \emph{support to autoscaling} (Figure~\ref{fig:monitoringAutomation} (c)), that is, probes automatically adapt to a changing number of replica of a  monitored service~\cite{trihinas2014jcatascopia}. This is a useful feature, although limited to a specific scenario, missing to cope with the many changes that must be actuated as a consequence of changes on the set of collected indicators and monitored services.
	
	To obtain a sufficient level of flexibility to address the aforementioned characteristics, \emph{Monitoring-as-a-Service} (MaaS) solutions have been recently studied~\cite{hp_2020_monasca,trihinas2014jcatascopia,FATEMA20142918,aceto2013monitoringsurvey} (Figure~\ref{fig:monitoringAutomation} (d)-(f)). 
	In fact, MaaS frameworks provide operators with the capability to flexibly decide the set of indicators to be collected, alleviating them from the burden of configuring and handling the life-cycle of the probes. In principle, an operator using a MaaS framework can simply specify the set of indicators that must be collected, while the operational aspects are automated by the framework. 
	
	Unfortunately, in many cases, automation is \emph{limited to the activation of manually pre-deployed probes}~\cite{trihinas2014jcatascopia} (Figure~\ref{fig:monitoringAutomation} (d)), that is, probes that have been already installed and configured \emph{manually}. Adding probes to collect new indicators and removing existing probes must still be done manually by operators.

	A higher degree of automation is provided by some \emph{specific platforms} (Figure~\ref{fig:monitoringAutomation} (e)) that natively offer monitoring capabilities (e.g., Kubernetes). These solutions are effective but significantly limit both the range of platforms and indicators that can be used.  \emph{So far, there is no general MaaS solution that can be used to collect virtually any KPI on any platform}. Note that a MaaS system that fully handles the life-cycle of probes is \emph{the only solution that can entirely free operators from the burden of handling probe deployment}; in fact they would be able to control the monitoring system by simply specifying the set of indicators to be collected.

	In this paper we address this challenge by presenting a MaaS monitoring framework (Figure~\ref{fig:monitoringAutomation} (f)) that exploits both a catalog of probes annotated with metadata and access to the API of the platform running the monitored services, to deliver \emph{full MaaS capabilities including error-handling}. 
	
	\changetxt{R1.C4}{This work extends our former tool demo paper~\cite{Tundo:Varys:ESECFSE:2019} by (i) proposing a consolidated and scalable architecture, (ii) introducing error handling capabilities in the monitoring system, (iii) providing a rigorous presentation of the monitoring framework, and (iv) reporting results from an extensive empirical evaluation of the effectiveness of the approach.} In a nutshell, this paper makes the following contributions:

	\added{
		\begin{itemize}[leftmargin=*]
			\item \textbf{Automated life-cycle management of probes}. We present a MaaS framework that fully automates the deployment and undeployment of arbitrary probes starting from declarative inputs (i.e., the list of indicators to be collected) entered by the operators. 
			\item \textbf{Scalable and independent control processes}. We \emph{rigorously} describe the probe deployment and undeployment procedures that guarantee the correctness of the resulting behavior. The involved processes are defined as stateless services to guarantee the scalability of the resulting system. 
			\item \textbf{\changetxt{R1.C2}{Deployment error} handling routines}. We present how errors in probe deployment can be autonomously detected and fixed by the MaaS framework. So far, \emph{error handling} capabilities received little attention, with approaches mostly focusing on error-free \changetxt{R1.C2}{deployment} scenarios or relying on the direct intervention of operators, despite the importance of error handling for long-running systems, such as a monitoring system~\cite{FATEMA20142918,aceto2013monitoringsurvey}. 
			\item \textbf{Technology-agnostic control processes}. We detail how the presented framework can be integrated with existing technologies (e.g., probe technology, backend database, and target environment) without the need of using ad-hoc solutions. 
			\item \textbf{Empirical evidence}. We empirically study the \emph{effectiveness} of the framework with both \emph{containers and virtual machines}, the \emph{efficiency of error-handling}, and the \emph{scalability} for an increasing number of requests. 
		\end{itemize}
	}
	
	The paper is organized as follows. Section~\ref{sec:runningExample} introduces a running example that is used to illustrate the approach throughout the paper. Section~\ref{sec:architecture} presents the MaaS framework that automates life-cycle management of probes, rigorously describing the control processes that govern probe deployment and undeployment. Section~\ref{sec:error} presents error handling. Section~\ref{sec:tech_agnostic} discusses support to frameworks and probe technologies. Section~\ref{sec:evaluation} presents empirical evidence. Section~\ref{sec:related_work} discusses related work. Finally, Section~\ref{sec:conclusions} provides concluding remarks.


	\section{Running Example} \label{sec:runningExample}
	In this section, we introduce a running example that we use to illustrate and exemplify how the proposed MaaS framework works. The example consists of a PostgreSQL instance {\sc target-PSQL} running as part of a larger cloud system. Such an instance is of interest for two operators: operator {\sc op-A} and operator {\sc op-B}. Operator {\sc op-A} is mostly interested in infrastructure KPIs and is collecting network consumption data related to {\sc target-PSQL}. Operator {\sc op-B} is interested in both infrastructure and application KPIs, and is collecting 3 KPIs: network consumption data, CPU consumption data, and database metrics. We refer to this initial configuration as {\sc init-conf}.
	
	In this context, operator {\sc op-A} may notice anomalous data in the network traffic and decide to collect information about two additional KPIs: CPU consumption and user session data. We refer to the configuration where operator {\sc op-A}  is also collecting these two additional KPIs as the {\sc 2-more-kpis-conf}.
	
	Finally, operator {\sc op-B} may loose interest for the PostgreSQL service, for instance because the services maintained by operator {\sc op-B} may stop using PostgreSQL. In such a case, operator {\sc op-B} stops collecting any indicator from {\sc target-PSQL}. We refer to this final configuration as  {\sc op-B-left-conf}.
	
	We will refer to these sample scenarios and configurations in the rest of the paper to explain how the set of probes necessary to collect the indicators required by operators  {\sc op-A} and  {\sc op-B} can be adjusted automatically and transparently to the operators. 
	

	\section{MaaS \deleted{Architecture}\added{Framework}}\label{sec:architecture}
	
	\begin{figure*}[ht]
		\centering
		\includegraphics[width=0.9\linewidth]{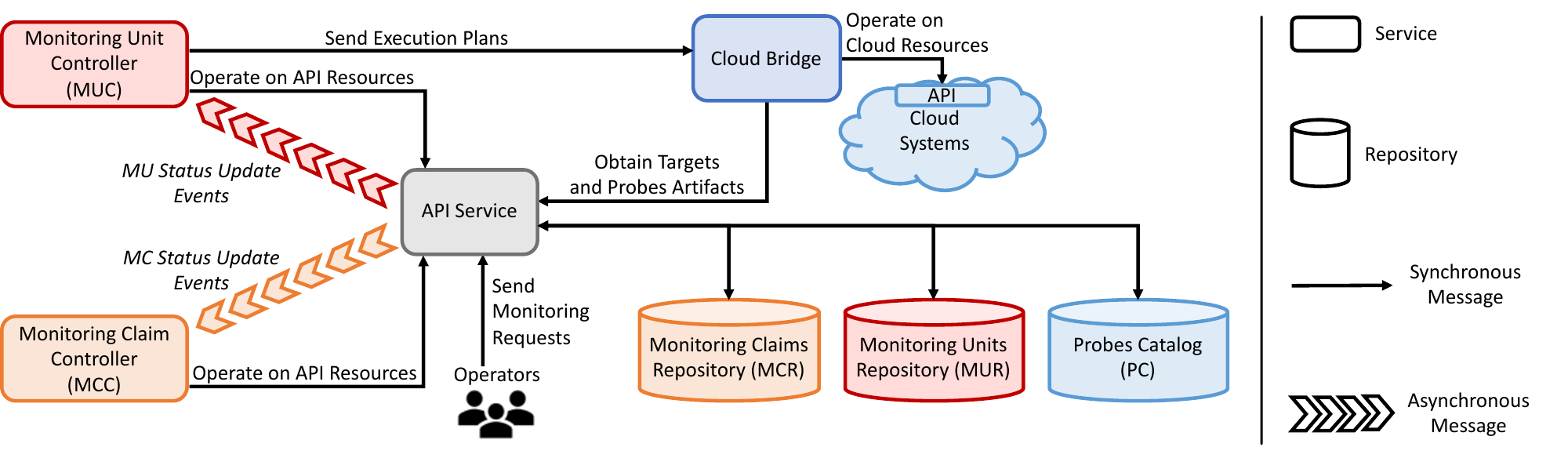}
		\caption{Monitoring \added{Framework} Architecture.}
		\vspace{-1.5em}
		\label{fig:architecture}
	\end{figure*}
	
	The proposed MaaS \deleted{architecture}\added{framework} exploits a few relevant domain concepts to organize the responsibilities of the components. In the following, we first introduce these key concepts, both informally and rigorously, and then discuss the \added{framework} architecture. 
	
	\smallskip
	A \mydef{Target} represents an application, a system, or a resource that can be monitored by a monitoring framework. It is characterized by the hosting platform (e.g., Microsoft Azure, Kubernetes), a unique identifier within the hosting platform (e.g., the Kubernetes Deployment name or VM name in Microsoft Azure), and its execution environment (e.g., virtual machine, or container). In our running example, the target is a PostgreSQL instance that we assume can be identified with the label {\sc target-PSQL} in both Kubernetes (as deployment name) and Microsoft Azure (as VM name).
	
	\smallskip

	A \mydef{Probe} represents a \emph{deployable artifact} that can be used to collect indicators from targets in different environments. Probes are annotated with metadata that describe how they can be deployed and configured.
	
	More rigorously, a probe $p$ is a tuple $p = (I, meta, artifact)$, where $I = \{i_1, \ldots i_n\}$ is a set of indicators that can be collected with the probe, 
	\changetxt{R2.C4}{$\textit{meta}$ is a set of key-value pairs that represent the metadata associated with the probe,} and $\textit{artifact}$ is a reference to the artifacts that implement the actual software probe. 
	We use the notation $p^I$, $p^{meta}$ and $p^{artifact}$ to refer to the individual components of a probe $p$.  
	
	\smallskip
	A \mydef{Monitoring Claim} specifies the indicators that an operator may want to collect for a specific target. 
	More rigorously, a monitoring claim $\textit{mc}$ is a tuple $\textit{mc}=(I, op, t)$ where $I=\{i_i, \ldots i_k\}$ is the set of indicators to be collected from the target $t$ for the operator $\textit{op}$. The claim is intended as a complete specification for the specified target, thus if the operator is already monitoring an indicator $i$ for a given target $t$ and the newly submitted claim does not include the indicator, the monitoring system will stop collecting $i$ from $t$. 
	For example, operator {\sc op-A} shall submit a monitoring claim $(\{\textit{\sc network\_consumption}, \textit{\sc cpu\_consumption}, \linebreak \textit{\sc user\_session\_data}\}, {\sc \textit{op-A}}, {\sc \textit{target-PSQL}})$ to start collecting CPU consumption and user session data, in addition to network consumption.  Similarly,  operator {\sc op-B} shall submit a monitoring claim $(\{\}, {\sc \textit{op-B}}, {\sc \textit{target-PSQL}})$ to stop collecting data.

	\smallskip
	A \mydef{Monitoring Request} is a collection of Monitoring Claims submitted with a single request by an operator. 
	More rigorously, a monitoring request $\textit{mr}$ submitted by operator $op$ is a set $\textit{mr}=\{\textit{mc}_1, \ldots \textit{mc}_m\}$ where $\textit{mc}_i=(I_i, op, t_i)$.
	
	\smallskip

	A \mydef{Monitoring Unit} is an execution unit (e.g., a virtual machine or a container) that runs one or more probes. When needed, our monitoring framework dynamically creates and destroys monitoring units to collect the indicators specified by the operators in their monitoring claims. A monitoring unit is also characterized by a hosting platform, which represents the environment where the unit is executed, and a configuration, which captures how the probes in the monitoring unit are configured.
	
	More rigorously, a monitoring unit $\textit{mu}$ is a tuple $mu  = (host, mus, C)$, where $host$ identifies the platform that provides the unit, $mus$ indicates the strategy used to configure the unit (i.e., single probe or multi-probe), and $C$ is the configuration of the unit, which consists of zero or more \emph{probe configurations}, depending on the number of probes installed. Specifically a probe configuration $c \in C$ is a tuple $ c=(p, I, op)$, where $p$ is a probe, $I \subseteq p^I$ represents the set of indicators that $p$ is configured to collect, and $op$ is the operator who asked for the probe configuration $c$.
	
	We use the notation $mu_P$ to refer to the set of probes in the current configuration of $mu$, that is, $mu_P= \{p | \exists (p, \cdot, \cdot) \in C\}$\footnote{We use the symbol $\cdot$ when any value is allowed in a tuple.}. Finally, given a probe configuration $(p, I, op)$, we use the notation $I(p)$ to refer to the indicators that $p$ is configured to monitor, that is, $I(p)=I$.
	
	Our framework implements two \textbf{monitoring unit configuration strategies}: the multi-probe monitoring unit and the single-probe monitoring unit. The \emph{multi-probe monitoring unit strategy} uses one monitoring unit (e.g., a virtual machine) per monitored target (e.g., an instance of PostgreSQL), hosting in the unit all  the probes that share a same target (e.g., every probe that collects indicators about PostgreSQL). This strategy is well suited for virtual machines, which are heavyweight units that typically run multiple processes. The \emph{single-probe monitoring unit strategy} uses one monitoring unit (e.g., a container) per deployed probe (e.g., a Metricbeat probe for CPU consumption). This strategy is well suited for containers, which are lightweight units that preferably run a single process.
	
	For instance, the initial configuration of the running example, where virtual machines running on Microsoft Azure are used, implies the existence of a single monitoring unit $\textit{mu}=(\textit{azure}, \textit{multi-probe}, C$), running the probe $p_{net}$, which serves both operators {\sc op-A} and {\sc op-B}, and the probes $p_{cpu}$, $p_{db}$, which both serve operator {\sc op-B}. Consequently, $C$ consists of the following four probe configurations: $(p_{net}, \textit{\sc network\_consumption}, \textit{\sc op-A})$, $(p_{net}, \textit{\sc network\_consumption}, \textit{\sc op-B})$, $(p_{cpu}, \textit{\sc cpu\_consumption}, \textit{\sc op-B})$, and $(p_{db}, \textit{\sc db\_metrics}, \text{\sc op-B})$.
	
	
	Note that the monitoring units are created to have the right visibility of the target to be monitored.
	In fact, a virtual machine monitoring unit can be either the same virtual machine running the monitored service or a separated virtual machine with probes that query an interface exposed by the monitored service (e.g., using SNMP~\cite{stallings1998snmp}).
	On the other hand, a container monitoring unit can be created as a sidecar of the container running the target service~\cite{burns2016design}, to have \added{extensive} visibility of the monitored \added{service}, \changetxt{R1.C1}{or as a standalone container running in the same node of the target.}
	
	\smallskip
	
	Figure~\ref{fig:architecture} shows the proposed \textbf{monitoring framework}, which consists of four main stateless services and three repositories.
	The four services are (i) an \emph{API Service}, which offers a gateway to access and update state information about the monitoring system, (ii) a \emph{Monitoring Claim Controller}, which is responsible for handling the life-cycle of every monitoring claim, (iii) a \emph{Monitoring Unit Controller}, which is responsible for handling the life-cycle of every monitoring unit, and (iv) a \emph{Cloud Bridge}, which exploits a plug-in based architecture to interact with different cloud providers and platforms, actuating the operations decided by the other services. The three repositories consist of (i) a repository of \emph{monitoring claims} submitted by operators, (ii) a repository with the created \emph{monitoring units} and their configurations, and (iii) a \emph{probe catalog} with all probes and deployable artifacts.  
	
	\emph{Automated life-cycle management} of the probes is provided by the two controllers that collaborate to manage the set of monitoring units, and the deployed probes, based on the requests produced by the operators that only include the information about the indicators to be collected. The stateless nature of the controllers guarantees \emph{scalability}, as long as sufficient resources are provided to the monitoring system. The controllers also track the status of the monitoring units to \emph{handle and recover from errors}. Finally, the framework is built with a plug-in based architecture that allows \emph{multiple cloud platforms} to be integrated, as long as they provide a management API. In the rest of this section, we rigorously describe how the components, and the controllers in particular, behave.

	\subsection{Repositories} \label{sec:datastores}
	The \mydef{Probe Catalog} is a repository $\textit{PC}=\{p_1,\ldots p_n\}$ where $p_i$ is a probe. We assume the probe catalog is organized in such a way there is a unique artifact that can be used in a given context, that is, given an index $i$ and the execution constraints (e.g., the host environment that executes the probe, the timeseries database that must be used to store the data, etc.), there is a unique probe $p$ 
	that can be used to collect $i$ in the target environment. We do not detail here the execution constraints that can be used to identify the probe, but these are represented in the metadata associated with the available artifacts and matched for equality (or inclusion in case of lists) by the framework to select the probes. 
	
	Complex matching procedures can be also implemented in the catalog if needed, such as the possibility to have multiple probes suitable for a same context, and a decision procedure that can choose among them. Defining algorithms to choose among multiple probe artifacts is however out of the scope of the presented work and we simply require the operator to populate the probe catalog with one usable artifact per execution context that must be addressed with the architecture.
	
	To illustrate the matching procedure, consider the case of {\sc op-A} asking to collect user session data from PostgreSQL. Let us assume the system considered in the running example runs on Kubernetes and that Elasticsearch is used as timeseries database. In this context, the monitoring system will check the probe catalog looking for a probe whose metadata specificy the capability to (a) collect user session data from PostgreSQL, (b) to run within containers, and (c) to store data in Elasticsearch. The monitoring system is configured with information about the environment (e.g., how to access Elasticsearch and Kubernetes APIs) to be able to configure the probes once deployed. If a matching entry is found, the corresponding artifacts are selected, and then deployed in a container, as illustrated later in this section. \changetxt{R2.C1}{Otherwise, the request is aborted and the Probe Catalog has to be extended to support new probes, as described in Section~\ref{sec:tech_agnostic}.} 
	
	
	The \mydef{Monitoring Claims Repository} stores the monitoring claims and tracks their statuses while they are created, processed, and updated. Since operators can update their claims about a given target, the repository can at most include one monitoring claim for a given operator-target pair.  For example, an operator may submit a first monitoring claim to collect network consumption for a running instance of PostgreSQL (corresponding to the {\sc init-conf} in our running example), and later update the monitoring claim asking to collect two more indicators, CPU consumption and user session data, still from PostgreSQL (corresponding to the {\sc 2-more-kpis-conf} in our running example).

	
	
	The \mydef{Monitoring Units Repository} tracks the status of the monitoring units and their configurations. In particular, the Monitoring Units Repository stores both the \emph{current configuration} of a monitoring unit, which reflects the status of the software monitoring unit, and the \emph{desired configuration} of a monitoring unit, which reflects the configuration that must be reached based on the received requests, supporting the controllers in the process of adapting the configurations. 
	
	To conveniently work with the configurations required by operators, we define the operator $|_{op}$ which discards every entry related to $op$ from a configuration. More formally, given a configuration $C$, we define $C|_{op}=\{c_i$ |  $c_i \in C$ and $c_i=(p_i, I_i, op_i)$ with $op_i \neq op$\}.

	A Monitoring Units Repository $\textit{MUR}$ stores tuples $(t, mu, dc)$ that associate a target $t$ with a monitoring unit $mu$ running probes that collect data from $t$, to its desired configuration $\textit{dc}$. 
	Given a monitoring unit $mu=(host, mus, C)$, we use the notation $conf_c(mu)$ to refer to its current configuration, that is,  $conf_c(mu)=C$. We instead use the notation $conf_d(mu)$ to refer to the desired configuration of a monitoring unit $\textit{mu}$, that is, $conf_d(mu)=dc$. 
	The level of alignment between $conf_c(mu)$ and $conf_d(mu)$ indicates how much the actual monitoring unit (i.e., the unit running in the cloud) matches  the monitoring claims submitted by operators. If $\textit{conf}_c(mu)=\textit{conf}_d(mu)$, the current and desired monitoring configurations are the same, thus the monitoring unit $\textit{mu}$ is up to date and perfectly aligned with the existing monitoring claims. Otherwise if $\textit{conf}_c(mu) \neq \textit{conf}_d(mu)$, the monitoring unit $\textit{mu}$ needs to be modified to reach the desired configuration.

	If $\textit{MUR}$ is handled according to the  multi-probe monitoring unit strategy, given a target $t$, there is at most one $mu$ such that $(t, mu, \cdot) \in \textit{MUR}$ (i.e., one monitoring unit running multiple probes per target). If $\textit{MUR}$ is handled according to the single-probe monitoring unit strategy, given a target $t$ and a probe $p$, there is at most one $(t, mu, C) \in \textit{MUR}$, with $(p, \cdot, \cdot) \in C$,  
	but there might exist multiple monitoring units running different probes associated with a same target.

	\subsection{API Service}
	The API Service provides two APIs: a public API for external clients and a private API for internal use only.
	
	The \emph{public API} is used by operators to submit monitoring requests, receive information about the status of their requests, extract the list of the current available Targets, and upload new probes to the Probes Catalog. 
	
	The \emph{private API} is used by the Monitoring Claim Controller and Monitoring Unit Controller to handle (i.e., to read and update) the status information about both the monitoring claims and the monitoring units, as described in Sections~\ref{sec:mcc} and~\ref{sec:muc}. 
	
	Note that the API Service is the only service that can directly access the three repositories. The presence of a single entry-point for accessing the persistent data drastically reduces the risk of (potentially) introducing data inconsistencies. To avoid introducing a single-point of failure in the architecture, we designed the API Service as a stateless service that can be instantiated in multiple replicas.
	
	The API Service is accessed through synchronous API calls, to guarantee that requests are processed as quickly as possible, but status updates are delivered through a message bus, since serving a request is not always an immediate operation.

	\subsection{Monitoring Claim Controller} \label{sec:mcc}
		%
		%
		%
		%
	\begin{algorithm}[ht]
		\small
		\caption{Monitoring Claim Controller}\label{alg:monitoring_claim_controller} 
		\begin{algorithmic}[1]
			\Require a monitoring claim $\textit{mc}=(I, op, t)$ to be processed 
			\Require $\textit{mus}$, the monitoring unit strategy
			\Ensure desired configurations are updated according to $mc$
			
			\item[]
			\State $P \gets$ APIService.getProbeConfigs($I,t$) \label{getProbesCall}
			\If{$P=\varnothing$} \Return \EndIf
			\item[]		
			\If{$\textit{mus}$=multi-probe}
			\State UpdateConfUnit($P$, $\textit{op}$, $t$, $mus$) \label{CallMultiProbe}
			\ElsIf{$\textit{mus}$=single-probe}
			\For{$p_{conf} \in P$} \label{forBegin}
			\State UpdateConfUnit($\{p_{conf}\}$, $\textit{op}$, $t$, $mus$) \label{forEnd}
			\EndFor
			\EndIf
			
			\item[]
			
			\Procedure{UpdateConfUnit}{Set of probe configurations $P$, operator $op$, target $t$, monitoring unit strategy $mus$}
			\State $\textit{unit} \gets$ APIService.getMonitoringUnit($t$, $mus$, $P$) \label{getMonitoringUnit}\label{searchMonitoringUnits}
			\If{$\textit{unit}=\varnothing$}
			\State $\textit{unit} \gets$ APIService.createEmpyMonitoringUnit($t$) \label{newUnit}
			\EndIf
			\State APIService.updateDesiredConf(unit, $conf_d(unit)|_{op} \cup P$) \label{replaceConfig}
			\EndProcedure
		\end{algorithmic}
	\end{algorithm}

	The main responsibility of the Monitoring Claim Controller is to manage the life-cycle of the submitted monitoring claims by assigning the desired configurations, derived from the received claims, with the monitoring units. 
	In particular, every time a monitoring request is received by the API Service, the API Service stores the monitoring claims included in the request in the dedicated repository and sends a status update message to the Monitoring Claim Controller, which will incrementally process them.
	
	Since controllers are stateless, the capability to process monitoring claims in parallel can be increased arbitrarily, based on the available resources, by instantiating multiple Monitoring Claim Controllers.
	
	Algorithm~\ref{alg:monitoring_claim_controller} shows in details the operations performed by the monitoring claim controller every time a monitoring claim is processed. When a monitoring claim $\textit{mc}=(I, op, t)$ of an operator $\textit{op}$ is processed, the controller first identifies the set of probes necessary to collect the indicators specified in the request and their configuration (line~\ref{getProbesCall}). This set is computed by the API service based on the probe metadata.
	
	The monitoring units are reconfigured differently depending on the monitoring strategy. If the \emph{multi-probe monitoring unit strategy is used}, the {\sc UpdateConfUnit} procedure is invoked to associate a single monitoring unit with a desired configuration that includes all the probes (line~\ref{CallMultiProbe}). If the \emph{single-probe monitoring unit strategy is used}, the individual probes configurations are extracted and then used to update the configuration of different monitoring units (lines~\ref{forBegin}-\ref{forEnd}).
	
	The way a set of probe configurations are associated with a monitoring unit is defined in the {\sc UpdateConfUnit} procedure. 
	To identify the monitoring unit that must be updated, the controller queries, through the API Service, the monitoring units repository for an existing monitoring unit (line~\ref{searchMonitoringUnits}). If the \emph{multi-probe monitoring unit strategy} is used, \changetxt{R3.C1}{units can conveniently run multiple probes for a same target}. In this case, the service looks for any monitoring unit created to observe $t$, that is, it looks for an entry $\textit{unit}=(t, \textit{multi-probe}, \cdot)$, where $t$ is the target reported in the monitoring claim. If the \emph{single-probe monitoring unit strategy} is used, \changetxt{R3.C1}{$P$ can only include a single probe}, and the API service looks for a monitoring unit that is already using the selected probe to monitor the target $t$, that is, it looks for an entry $unit  = (t, \textit{single-probe}, (p, \cdot, \cdot) )$.
	
	In both cases, if the unit does not exist, a new unit with an empty desired configuration is created for the target $t$ (line~\ref{newUnit}). Finally, the existing entry (i.e., the existing desired configuration) is updated by replacing the probes associated with operator $op$ with the new ones specified in $P$ (if the existing configuration is empty, $P$ is simply used).
	
	Let us consider the running example, with operator {\sc op-A} asking to collect two more indicators (CPU consumption and user session data) from PostgreSQL, if we assume the monitoring framework is configured to use the single-probe monitoring strategy, the submitted monitoring claim would be processed as follows. The access to the probe metadata would reveal the availability of two different probes that can be configured to collect the two indicators: $p_{cpu}$, which can monitor CPU consumption using a Metricbeat probe, and $p_{session}$, which can use a custom probe to collect data about user sessions. That is, $P$=\{($p_{cpu}$, {\sc cpu\_consumption}, {\sc op-A}), ($p_{session}$, {\sc user\_session\_data}, {\sc op-A})\} at line~\ref{getProbesCall}. Since $mus$=$\textit{single-probe}$, the {\sc updateConfUnit} procedure is invoked twice, once for each probe.
	
	The first invocation with probe $p_{cpu}$ leads to the identification of a running unit that is already collecting {\sc cpu\_consumption} from PostgreSQL for  {\sc op-B} (line~\ref{getMonitoringUnit}). The current configuration of the retrieved unit is \{$p_{cpu}$, {\sc cpu\_consumption}, {\sc op-B})\}. The framework finally updates the desired configuration of the unit by replacing the probe configurations of operator {\sc op-A} (none in this case) with the input configuration ($p_{cpu}$, {\sc cpu\_consumption}, {\sc op-A} ), finally obtaining the desired configuration  \{($p_{cpu}$, {\sc cpu\_consumption}, {\sc op-B}), ($p_{cpu}$, {\sc cpu\_consumption}, {\sc op-A})\}. 
	
	The second invocation with probe $p_{session}$ returns no unit that is already running that probe. Thus, a new unit is created (line~\ref{newUnit}), and the desired configuration \{($p_{session}$, {\sc user\_session\_data}, {\sc op-A})\} is associated with the unit.
	
	\changetxt{R2.C2}{The time complexity of Algorithm~\ref{alg:monitoring_claim_controller} is linear with respect to the number of selected indicators ($I$) and the number of matched probes ($P$), that is, $O(|I| + |P|)$.}
	
	\subsection{Monitoring Unit Controller} \label{sec:muc}
	
	\begin{algorithm}[ht]
		\caption{Monitoring Unit Controller}\label{alg:monitoring_unit_controller} \label{alg:muc} \small
		\begin{algorithmic}[1]
			\Require a monitoring unit $mu$
			\Require its current configuration $\textit{conf}_c(mu)=\{(p, I, \textit{op})\}$
			\Require its desired configuration $\textit{conf}_d(mu)=\{(p', I', \textit{op}')\}$
			\Ensure the unit is updated according to the desired configuration is generated
			
			\If{$\textit{conf}_d(mu)=\varnothing$} dismiss $mu$ \label{dismissUnit}
			\EndIf

			\State $P_{add} \gets \{p \in \textit{conf}_d(mu)_P\setminus \textit{conf}_c(mu)_P \}$ \label{probesAdd}
			\State $P_{update} \gets \{p \in \textit{conf}_d(mu)_P \cap  \textit{conf}_c(mu)_P$ s.t. $ I'(p) \not= I(p) \}$ \label{probesUpdate}
			\State $P_{drop} \gets \{p \in \textit{conf}_c(mu)_P \setminus \textit{conf}_d(mu)_P \}$ \label{probesDrop}
			
			\If{$P_{add} \cup P_{update} \cup P_{drop} \neq \varnothing$} 
			
			\State $\textit{res} \gets$ Bridge.doChanges($mu$, $P_{add}, P_{update}, P_{drop}$) \label{bridge}
			\Else
			\State $\textit{res} \gets \varnothing$
			
			\EndIf

			\State UpdateConfiguration($mu$, res) \Comment{{\footnotesize If no error,  $\textit{conf}_c(mu)$ is updated with $\textit{conf}_d(mu)$}} \label{updatedOperation}
		\end{algorithmic}
	\end{algorithm}
	
	The main responsibility of the Monitoring Unit Controller is to manage the life-cycle of the monitoring units according to the desired configurations generated by the Monitoring Claim Controller. 
	In particular, the Monitoring Unit Controller runs a control-loop that 
	continuously checks the Monitoring Units for changes to be actuated, as a consequence of a misalignment between the current and the desired configurations. Multiple monitoring unit controllers can be active at the same time, but two monitoring unit controllers cannot act simultaneously on a same monitoring unit, \changetxt{R3.C1}{to prevent any potentially erroneous concurrent change that would introduce inconsistencies in the process}.
	
	The operations performed by a Monitoring Unit Controller are shown in Algorithm~\ref{alg:muc}. It first checks if the desired configuration is empty, in such a case the entire monitoring unit is dismissed (line~\ref{dismissUnit}). This is an important step to avoid running phantom monitoring units with no running probes. It then computes the diff between the current and desired configuration, identifying the probes to be added (line~\ref{probesAdd}), the probes to be reconfigured to collect a different set of indicators (line~\ref{probesUpdate}), and the probes to be dropped (line~\ref{probesDrop}). If any of these sets is non empty, the Cloud Bridge receives the probe configurations corresponding to the changes that must be actuated (line~\ref{bridge}). \changetxt{R3.C1}{Passing all the changes to be actuated at once enables the Cloud Bridge to potentially optimize how these changes are actuated.}
	
	The Cloud Bridge returns a result that specifies the errors experienced during the update process, if any. This information is used to update the current and desired configuration. In case no error is experienced, the desired configuration simply replaces the current configuration (line~\ref{updatedOperation}). Otherwise, the update process takes the errors into consideration. We describe error handling in Section~\ref{sec:error}.
	
	Let us consider the case of the two desired configurations generated by operator {\sc op-A} when asking to collect two more indicators (CPU consumption and user session data) from PostgreSQL with the single-probe monitoring unit strategy, as discussed at the end of Section~\ref{sec:mcc}. The desired configuration related to the already deployed probe $p_{cpu}$ results in no changes to be operated ($P_{add} \cup P_{update} \cup P_{drop} = \varnothing$), since the existing probe will be simply shared between the two operators (this is achieved by only updating the configurations in {\sc UpdateConfiguration} without touching the running probes). While, the desired configuration related to the new probe $p_{session}$ to be deployed results in a probe to be added ($P_{add} \neq \varnothing$). 
	
	\changetxt{R2.C2}{The time complexity of Algorithm~\ref{alg:monitoring_unit_controller} is linear with respect to the number of probes to add ($|P_{\textit{add}}|$), update ($|P_{\textit{update}}|$), and drop ($|P_{\textit{drop}}|$) while configuring a monitoring unit. That is, if $\textit{pchanges}=|P_{\textit{add}}| + |P_{\textit{update}}| + |P_{\textit{drop}}|$,  the complexity of Algorithm~\ref{alg:monitoring_unit_controller}  is $O(\textit{pchanges})$.}

	\subsection{Cloud Bridge}
	
	The main responsibility of the Cloud Bridge is to actuate plans on cloud systems using their management APIs. The Cloud Bridge also provides information about the targets and the deployment status of the probe artifacts.
	
	In particular, the Cloud Bridge exploits a plug-in based architecture that can be extended to support additional cloud systems. A plug-in for a target environment (e.g., Kubernetes) is used to map each change requested by controllers into a concrete command for the specific management API  (e.g., the Kubernetes API) or the specific configuration management tool used to interact with the platform (e.g., Ansible~\cite{redhat_2020_ansible}). This approach encapsulates the technological details inside the plug-in, keeping the whole control-plane framework agnostic from technology. Once all the changes have been actuated, the list of probes resulting in an erroneous state is sent back to the controller. 
	
	
	
	
	\section{Error Handling}\label{sec:error}
	
	
	\changetxt{R1.C2}{The presented \deleted{architecture}\added{framework} implements error handling procedures to recover from deployment errors, namely, errors that might be experienced at deployment time while creating, updating and removing either probes or monitoring units. The framework does not target the runtime errors that might be experienced after a successful deployment.}
	These procedures are extremely important for the dependability of the monitoring framework, whose behavior may otherwise diverge from the desired behavior. We distinguish two classes of errors that can be detected and handled:
	\begin{itemize}[leftmargin=*]
		\item \textbf{Soft errors}. Soft errors indicate problems in the operations performed \emph{while preparing} for the creation, update and deletion of a unit, such as retrieving probes and preparing their configuration. All these operations are performed \emph{before} modifying any existing monitoring unit. Since those are problems that do not compromise the dependability of the running units, they are considered soft errors that have negligible consequences on the running monitoring system.  
		\item \textbf{Hard errors}. Hard errors indicate problems in the operations performed while \emph{changing a running monitoring unit}, such as adding, reconfiguring or removing probes. Since these problems may compromise the dependability of the running monitoring system, they are considered hard errors that timely require corrective actions to be managed.  
	\end{itemize}
	
	Errors are detected by the Cloud Bridge while interacting with platform management APIs and while running commands of configuration systems. Soft errors are produced during the execution of the preparatory steps, differently from hard errors that are generated while changing the actual monitoring units. For this reason, depending on if and when an error is detected, a probe to be deployed can be in one of the following states:
	\begin{itemize}[leftmargin=*]
		\item \emph{Failed probe}: a soft error has been detected by the Cloud Bridge while preparing the probe.
		\item \emph{Broken probe}: a hard error has been detected by the Cloud Bridge while deploying/undeploying the probe.
		\item \emph{Stable probe}: no error detected
	\end{itemize}
	
	The errors detected for each probe configuration that is processed by the Cloud Bridge are reported in the results returned to the Monitoring Unit Controller (line~\ref{bridge} of Algorithm~\ref{alg:muc}).
	
	Consequently, a monitoring unit can be in any of the following states, depending on the states of its probes:
	\begin{itemize}[leftmargin=*]
		\item \emph{Stable unit}: no error is detected for the probes in the monitoring unit.
		\item \emph{Unsound unit}: there is at least a failed probe and no broken probe in the monitoring unit. This status indicates a failure in the attempt to align the desired and current configurations of the monitoring unit, but no actual problem is affecting the running unit. 
		\item \emph{Dirty unit}: there is at least a broken probe in the monitoring unit. This status indicates that the software running in the unit might be  compromised.
	\end{itemize}
	
	\begin{algorithm}[ht]
		\caption{UpdateConfiguration}\label{alg:error_handling} \label{alg:errorHandling} \small
		\begin{algorithmic}[1]
			\Require a monitoring unit $mu$ to be updated
			\Require $res=(Pconf_{soft}, Pconf_{hard})$, where $Pconf_{soft}$ and $Pconf_{hard}$ are the set of probe configurations that resulted in soft or hard errors  
			\Require $RetryTable \subseteq MUnits \times ProbeConfigs \times \mathbb{N}$, which is a table that counts how many times a given probe configuration has been retried in a monitoring unit
			\Require $BlackList \subseteq MUnits \times ProbeConfigs$, which is a table that tracks the probe configurations that cause errors and should not be retried again
			
			\Ensure $mu$ is updated and any error is reported 
			
			\For{$pc \in Pconf_{soft}$}
			\State RetryTable.IncRetry(mu, pc) \label{alg:retry}
			\EndFor		
			
			\For{$pc \in Pconf_{hard}$}
			\State BlackList.add(mu, pc) \label{alg:blacklist}
			\EndFor		
			
			\If{$Pconf_{hard} \neq \varnothing$} \Comment{{\footnotesize Dirty unit}}
			\State Bridge.cleanUnit(mu) \label{cleanUnit}
			\State $conf_c(mu) \gets \varnothing$
			\Else
			\State $conf_c(mu) \gets conf_d(mu) \setminus (Pconf_{soft} \cup Pconf_{hard})$ \Comment{{\footnotesize $conf_d(mu)$ is unchanged, so probe configs causing soft errors are retried, while probe configs with too many retries and probe configs in blacklist are automatically ignored}}
			\EndIf
			
		\end{algorithmic}
	\end{algorithm}
	
	Errors are mostly handled in the context of the {\sc UpdateConfiguration} procedure whose pseudocode is shown in Algorithm~\ref{alg:errorHandling}. The {\sc UpdateConfiguration} procedure is invoked by the Monitoring Unit Controller to finalize the update of a monitoring unit (line~\ref{updatedOperation} in Algorithm~\ref{alg:muc}). 
	
	In addition to referring to a monitoring unit $mu$ and the set of probe configurations that resulted in soft ($Pconf_{soft}$) and hard ($Pconf_{soft}$) errors, the procedure maintains two data structures. The $RetryTable$ is a table that stores for every monitoring unit the number of consecutive soft failures generated by each probe configuration. The $BlackList$ data structure stores for each monitoring unit the list of probe configurations that generated hard failures. The idea is that soft failures are not harmful for the monitoring unit, and thus the failed changes can be safely retried. Instead, hard failures introduce dependability problems, and thus the failed changes should  not be retried. Operators can reset these tables to allow again certain operations (e.g., after a compatibility problem in a probe has been fixed). 
	
	In practice, the error handling routine first increases the number of retries for the probe configurations that caused soft failures (line~\ref{alg:retry}) and adds to the blacklist the probe configurations that caused hard errors (line~\ref{alg:blacklist}). When the number of retries exceeds an operator-defined threshold, the configuration is blacklisted.  
	
	If at least a hard error has been detected, the unit is \emph{dirty} and thus the bridge is asked to clean it. This operation depends on the target environment and the implementation of the plug-in used in the Cloud Bridge. For instance, in our implementation for containers, the bridge destroys the existing container and creates a new monitoring unit to replace it. The current configuration of the newly created monitoring unit is consequently set to the empty configuration.
	
	If no hard error is detected, the current configuration is updated by adding all the configurations that generated no errors. In all the cases, the desired configuration stays unchanged. 
	
	This process may lead to three main distinct situations:
	\begin{itemize}
		\item \emph{the current and desired configurations are aligned}: no changes will be performed on the monitoring unit in the future, unless a new request is submitted by an operator;
		\item \emph{the current and desired configuration differs only for some blacklisted configurations}: in this case again there is nothing to be done. Note that although for simplicity we have not used the blacklist when computing the set of probes to be added, reconfigured, and deleted, in reality the Monitoring Unit Controller discards the configurations that appear in the BlackList data structure when computing them (Algorithm 2, lines~\ref{probesAdd}-~\ref{probesDrop}) 
		\item \emph{there are configurations that must be retried}: in such a case the desired and current configurations do not match, and the monitoring unit controller will process them again in the next iteration of its control-loop, retrying the failed probe configurations.
	\end{itemize}
	
	\changetxt{R2.C2}{The time complexity of the Algorithm~\ref{alg:error_handling} is linear with respect to the number of probe changes and number of errors occurred while configuring the monitoring unit. In particular, if $\textit{errors}$ is the number of probe configurations that resulted in soft or hard errors. The resulting time complexity is $O(\textit{pchanges} + \textit{errors})$.}

	\section{Technology Agnostic Design}\label{sec:tech_agnostic}
	The proposed monitoring framework is designed to transparently integrate heterogeneous monitoring technologies, releasing a \emph{technology agnostic control-plane} that can be exploited to obtain MaaS capabilities using the preferred probe technologies and target platforms. To witness this capability, this section exemplifies the integration of probes of different types and the capability to support multiple cloud platforms. 
	
	\begin{figure*}[!t]
		\begin{minipage}[t]{0.3\linewidth}
			\centering
			\begin{minted}[obeytabs=true, tabsize=1, fontsize=\scriptsize]{json}
				{
					"id": "5fb6337a4102891e3677b475",
					"artifactId": "kafka_exporter",
					"supportedIndicators": [
					"KAFKA_BROKERS", "..."
					],
					"supportedDataOutputs": [
					"PROMETHEUS"
					],
					"supportedMUStrategies": [
					"SINGLE_PROBE",
					"MULTI_PROBE"
					]
				}
			\end{minted}
			\subcaption{}
			\label{lst:kafka_probe}
		\end{minipage}
		\begin{minipage}[t]{0.33\linewidth}
			\centering
			\begin{minted}[obeytabs=true, tabsize=1, fontsize=\scriptsize]{json}
				{
					"id": "5fb6337a4102891e3677b476",
					"artifactId": "http_healthcheck_probe",
					"supportedIndicators": [
					"HEALTHCHECK"
					],
					"supportedDataOutputs": [
					"ELASTICSEARCH"
					],
					"supportedMUStrategies": [
					"SINGLE_PROBE",
					"MULTI_PROBE"
					]
				}
			\end{minted}
			\subcaption{}
			\label{lst:http_probe}
		\end{minipage}
		\begin{minipage}[t]{0.33\linewidth}
			\begin{minted}[obeytabs=true, tabsize=1, fontsize=\scriptsize]{json}
				{
					"targetPlatform": "azure",
					"targetPlatformId": "postgres-1",
					"envType": "INACCESSIBLE_VM",
					"metadata": {
						"resourceGroup": "resource-group-vm-1",
						"ipAddress": "52.92.34.124",
						"privateIpAddress": "10.19.20.3",
						"..."
					}
				}
				
				
				
			\end{minted}
			\subcaption{}
			\label{lst:target}
		\end{minipage}
		\captionof{listing}{Listings show an excerpt of metadata for the Kafka Prometheus Exporter (a), an excerpt of metadata for the HTTP Healthcheck Probe (b), and a sample JSON representation of a Target retrieved from Microsoft Azure (c).}\label{lst:metadata}
		\vspace*{-1.0em}
	\end{figure*}
	
	\subsection{Incorporating New Probes}
	%
	
	To demonstrate the flexibility of the monitoring framework we describe how two largely different probes can be supported: a health-check probe, which queries the health status endpoint of services exploiting the HTTP protocol, and a Prometheus exporter for Apache Kafka~\cite{apache_2020_kafka}, which monitors Kafka brokers resources (topics, partitions, etc.) and exposes the collected indicators as Prometheus metrics.
	
	\changetxt{R2.C1}{Adding a new probe can be done in two steps. First, the probe artifacts have to be manipulated in such a way they can be used by the Cloud Bridge. Second, the probe is added to the catalog by passing the probe metadata, which include information about where the probe can be deployed (the probe might be compatible with certain monitoring unit strategies but not with others), the supported data outputs (i.e., the database where the collected values can be stored), and the supported indicators, to the API Service.}
	Listings \ref{lst:metadata} (a) and (b) show an excerpt of the metadata associated with the Apache Kafka Prometheus exporter and the health-check probe, respectively. Note that the configuration of the monitoring framework (e.g., the knowledge of both the available time-series database and the type of the target platform), jointly with the requests produced by the operators, allows the framework to select and deploy the right probes. In fact, artifact ids are mapped to the concrete software artifacts and scripts that are executed for probe deployment. 
	
	Adding new probes (i.e., new artifacts and corresponding metadata) to the catalog may require a different amount of time depending on the knowledge of the involved technologies. It is however a quite convenient operation for people who know the monitoring framework. For instance, we needed 1.5 hours to develop and setup a health-check probe that can be deployed on virtual machines, and 30 minutes to add a Kafka exporter that can be deployed on Kubernetes.       
	
	\subsection{Supporting New Target Cloud Platforms}
	
	
	Supporting multiple target cloud platforms is another capability of the framework. A platform can be supported only if it provides a management API that can be used by the Cloud Bridge to manage the monitoring units and discover targets. Developers who want to create a new Cloud Bridge plug-in have to implement the base interface in order to run execution plans and provide information about the targets to the framework. Listing \ref{lst:metadata} (c) shows an example of target information that can be retrieved by the API via the Cloud Bridge component. Plug-ins are also associated with metadata (e.g., the supported monitoring unit strategies) that can help the framework in taking some decisions.

	
	Our prototype implementation already includes two plug-in implementations that can transparently actuate the same plans on  radically different platforms: Kubernetes, a container-based platform, and Microsoft Azure Compute, a virtual-machine-based platform. 

	\section{Empirical Evaluation}\label{sec:evaluation}
	Since probe deployment and error handling are two representative capabilities of the proposed framework, we designed an empirical evaluation to assess them. We further study scalability, to investigate how the monitoring framework scales with an increasing number of requests and operators. These points are captured by the following three research questions.
	
	\deleted{
		\begin{enumerate}[leftmargin=*]
			\item \textbf{RQ1 - How efficiently are probes deployed?} This research question investigates how efficiently the architecture can be used to deploy probes. This is investigated for both cloud systems based on containers and virtual machines. Results are studied in comparison to architectures working with pre-deployed probes. \alessandro{Forse dovremmo subito dire JCatascopia? Dato che alla fine è solo 1}
			
			\item \textbf{RQ2 - How efficiently are errors handled?} This research question investigates the time required by the architecture to detect and recover from errors. This is studied for both cloud systems based on containers and virtual machines. 
			
			\item \textbf{RQ3 - How does the architecture scale for an increasing number of requests?} This research question studies scalability with respect to the number of requests produced by operators who may ask for the same or different indicators. 
		\end{enumerate}
	}
	
	\added{
		\begin{enumerate}[leftmargin=*]
			\item \textbf{RQ1 - \emph{Framework Efficiency}: How efficiently are probes deployed?} This research question validates the framework capability of deploying probes starting from a declarative input and investigates how efficiently it is in fulfilling an operator monitoring request. This is investigated for both cloud systems based on containers and virtual machines giving evidence of the technology-agnostic capabilities of the framework. Results are studied in comparison to \changetxt{R2.C5}{a solution working with pre-deployed probes that can be activated/deactivated (Figure~\ref{fig:monitoringAutomation}, cases (d) and (e)). To this end, we selected JCatascopia~\cite{trihinas2014jcatascopia}, which is consistent with the MaaS case shown in Figure~\ref{fig:monitoringAutomation} (d), and it is usable with no restrictions being an open-source research prototype.} 
			
			
			\item \textbf{RQ2 - \emph{Error Handling}: How efficiently are errors handled?} This research question validates the framework capability of detecting and recovering from errors and investigates the time required by the framework to execute the error handling routine. 
			
			\item \textbf{RQ3 - \emph{Scalability}: How does the framework scale for an increasing number of requests?} This research question validates the framework capability of optimizing the control-plane during the evolution of the monitoring system. It studies scalability with respect to the number of requests produced by operators. 
		\end{enumerate}
	}
	
	All RQs were addressed with cloud systems based on both virtual machines and containers. In the following, we describe the prototype we used to run experiments, we report the design of the study, and the results for each research question.
	
	\subsection{Prototype} \label{sec:prototype}
	We implemented the framework described in this paper in a publicly available prototype hosted at \url{https://gitlab.com/learnERC/varys}.
	
	The services are implemented as Java standalone applications. The repositories are implemented as MongoDB~\cite{mongodb_2020_mongodb} collections. The JSON format is used both for communication and to persist information, except for the Cloud Bridge which exposes a gRPC API that uses Protocol Buffers. The status update messages are delivered through Redis Streams~\cite{redis.io_2020_redis}. The monitoring system can be deployed both on containers and virtual machines, depending on the hosting environment. 
	
	We designed a probe catalog reusing probes from Metricbeat~\cite{elasticsearch_2020_beats}, one of the most popular cloud monitoring framework. We used Elasticsearch~\cite{elastic_2020_elasticsearch}, as timeseries database to store the values extracted by the probes. We implemented plug-ins for the Cloud Bridge to support both Kubernetes and Microsoft Azure as target cloud platforms.
	
	\changetxt{R1.C1}{The Microsoft Azure plug-in supports either creating virtual machine monitoring units on-the-fly within the configured Azure resource group, or accessing the same virtual machine running the target to deploy the probes internally. In our experiments, we annotated the target service as an \textsc{ACCESSIBLE\_VM}, and made it accessible to the Cloud Bridge via SSH in order to (un)deploy the probes directly within the virtual machine running the target service.
		
		With respect to container monitoring units, the Kubernetes plug-in deploys container monitoring units in the same platform of the target and configures the probes accordingly. Purposely, it does not implement the container sidecar pattern~\cite{burns2016design} because it would trigger the redeployment of the target service, due to how Pods work in Kubernetes, every time probes are (un)deployed, potentially causing service or monitoring interruptions unless a robust rolling update strategy is in place.}
	
	\subsection{RQ1 - How efficiently are probes deployed?}
	\begin{figure}[ht]
		\begin{center}
			\begin{subfigure}[b]{0.4\textwidth}
				\includegraphics[width=\textwidth]{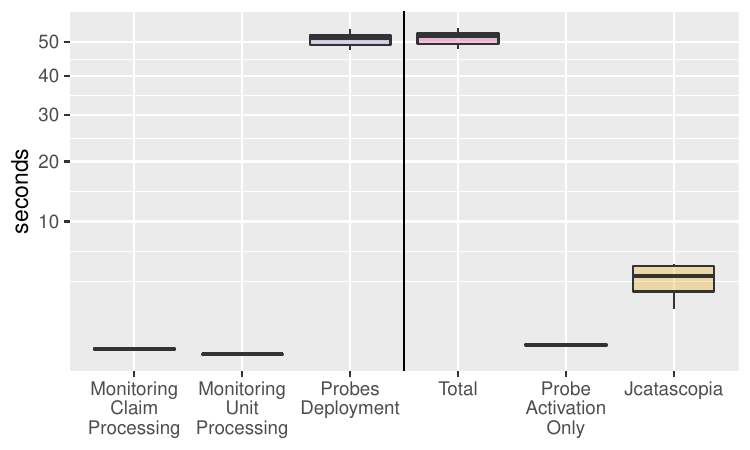}
				\caption{Virtual Machines Running on Azure}
				\label{subfig:azure}
			\end{subfigure}\\
			\begin{subfigure}[b]{0.4\textwidth}
				\centering
				\includegraphics[width=\textwidth]{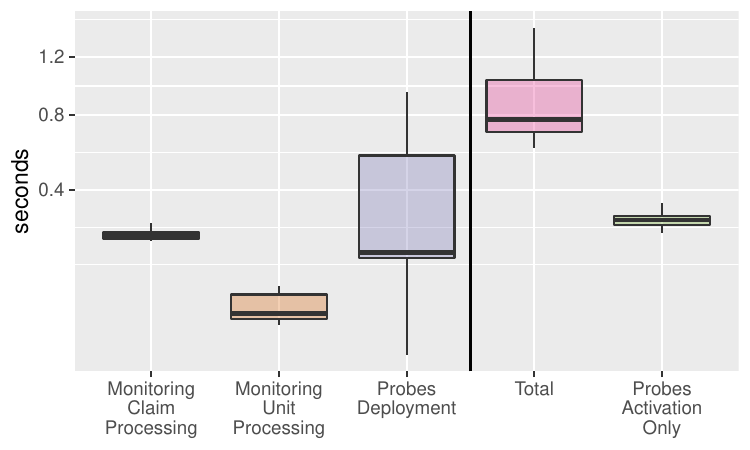}
				\caption{Containers Running on Kubernetes}
				\label{subfig:kubernetes}
			\end{subfigure}
			\caption{Probes Deployment}
			\label{fig:probes_deployment}
		\end{center}
		\vspace{-1.5em}
	\end{figure}

	The monitoring \deleted{architecture}\added{framework} can work in parallel on any number of monitored targets, if enough instances of the monitoring unit controller service are created. If there are more targets to modify than controller instances, some modifications will be performed sequentially. For instance, if 4 monitoring units must be modified and only 3 controller instances are available, one unit will be modified sequentially after another one. We will thus study how efficiently a monitoring unit can be managed by a single controller instance, the performance over multiple simultaneously evolving units can be straightforwardly deduced given the number of controllers available.
	
	We consider two cases for the experiments: PostgreSQL running in a container in an on-premise installation of Kubernetes and PostgreSQL running in a virtual machine on Microsoft Azure. The two cases show how the same framework can be transparently used to address heterogeneous scenarios where the involved technologies are significantly different.  We collect time figures considering the case of a request that requires the simultaneous deployment of three probes to collect the following three indicators from PostgreSQL: CPU consumption (using the CPU metricset of the Metricbeat probe), memory consumption (using the memory metricset of the Metricbeat probe), and database metrics (using the database metricset of the Metricbeat probe).
	
	To study the efficiency of each step, we measure the time taken by the first controller to process the claim, by the second controller to compute the execution plan, and by the Cloud Bridge to actuate the plan. 
	To have a baseline measurement, we also consider the case of a static \deleted{architecture}\added{framework}, that is, a \deleted{architecture}\added{framework} that does not support dynamic probe deployment, requiring operators to \emph{\deleted{manually} deploy and configure probes in-advance}, which can be later activated and de-activated. This \deleted{architecture}\added{framework} is \emph{far less flexible} than the \deleted{architecture}\added{framework} presented in this paper, but faster since it does not deploy probes dynamically. To this end, we both use our \deleted{architecture}\added{framework} with pre-deployed probes and the JCatascopia~\cite{trihinas2014jcatascopia} state of the art monitoring solution, which allows us to collect further measurements from a third party system. We do not have measurements for JCatascopia applied to containers since it only supports virtual machines. Every experiment is repeated 10 times to collect stable measures.
	
	Figure~\ref{fig:probes_deployment} shows the collected time figures with a semilogarithmic scale 
	considering both virtual machines (Figure~\ref{subfig:azure}) and containers (Figure~\ref{subfig:kubernetes}). The individual steps of the probe deployment process are captured by the Monitoring Claim Processing, Monitoring Unit Processing and Probes Deployment boxes. While Total represents the total time elapsed between the submission of the request and the time the deployed probes start collecting the required indicators. 
	
	Not surprisingly Probes Deployment is the most expensive step of the process \changetxt{R3.C3}{for both virtual machines and containers}. In the case of virtual machines it takes nearly 50 seconds, while the other steps can be completed an order of magnitude faster. In case of containers the difference is remarkably smaller, due to their computational efficiency and their ability to cache artifacts. In fact, probes deployment can be performed in at most 1 second with containers, while the remaining steps take less than 0.25 seconds. 
	
	Overall, the entire probe deployment process \changetxt{R3.C3}{of the three probes (indicated with Total in Figure~\ref{fig:probes_deployment})} could be completed in slightly less than a minute using virtual machines and less than 1.5s using containers, which is a nearly two orders of magnitude difference. 
	
	The box Probe Activation Only shows the time required to activate pre-deployed probes using our \deleted{architecture}\added{framework}. In the case of virtual machines, exploiting dynamic probe deployment might be quite expensive compared to manually pre-deploying probes, since it increases the runtime cost by an order of magnitude. 
	However, pre-deploying many probes can be expensive, can generate large and difficult to manage virtual machines, and is efficient only when the indicators that might be collected can be predicted. 
	The comparison to JCatascopia shows that the presented \deleted{architecture}\added{framework} is efficient, also when just used to process requests and activate pre-deployed probes. In fact, JCatascopia required several seconds to activate the probes, while our \deleted{architecture}\added{framework} could activate probes in less than a second. \changetxt{R3.C3}{The difference between dynamic probe deployment and pre-deployed probes for containers is indeed less significant, both in relative and especially absolute terms.}
	
	\smallskip
	
	\textbf{Answer to RQ1} In the case of virtual machines, the cost of flexibly deploying probes is significantly higher than working with pre-deployed probes. Thus, the trade-off between flexibility and efficiency should be carefully considered by operators to decide the monitoring solution that should be adopted. 
	Instead, in the case of containers, the cost of flexibly deploying probes is significantly amortized by the efficiency of the cloud technology. In fact, our \deleted{architecture}\added{framework} can complete the process in 0.5-1.5 seconds, while activating the pre-deployed probes requires slightly less than 0.5s, suggesting that dynamic probe deployment might be often preferable.

	\subsection{RQ2 - How efficiently are errors handled?}
	
	To study the capability of the \deleted{architecture}\added{framework} to react to errors, we designed a variant of the experiment performed for RQ1 where we deploy a malfunctioning probe. We obtained such a probe by implementing a wrong configuration of the Metricbeat probe for PostgreSQL \changetxt{R1.C2}{that makes the probe deployment to fail}. 
	
	In the case of virtual machines, we study the creation of a new monitoring unit with two probes: one working probe and a malfunctioning probe. \changetxt{R1.C2}{The malfunctioning probe artifact contains an Ansible role with a wrong command that leads to a hard deployment error when the Cloud Bridge executes it.} Since we use the multi-probe monitoring unit strategy with virtual machines, error detection must autonomously detect the problem with the monitoring unit with two probes and automatically create a monitoring unit with the working probe only.
	
	In the case of containers, we study the creation of a new monitoring unit with the malfunctioning probe only. \changetxt{R1.C2}{The malfunctioning probe artifact contains a bugged Kubernetes manifest file that tries to deploy the probe within a non-existent Kubernetes namespace. This leads to a hard deployment error when the Cloud Bridge executes it.} Since we use the single-probe monitoring unit strategy with containers, error detection should simply drop the malfunctioning monitoring unit (in this case we do not consider the deployment of two probes because the deployment strategy would simply create two different monitoring units handled independently).
	
	To capture how error detection works, we measure the time necessary to the \deleted{architecture}\added{framework} to \emph{attempt the deployment and detect} that a monitoring unit is not working (namely Error Detection), the time necessary to \emph{process} the error and take the decision to clean the monitoring unit (namely Error Processing), and finally the time necessary to actuate the cleaning plan (namely Error Cleaning). Error detection is performed by the cloud bridge while actuating changes (see the call in Algorithm~\ref{alg:muc}, line~\ref{bridge}), error processing consists of the operations shown in Algorithm~\ref{alg:errorHandling}, and error cleaning is again performed by the Cloud Bridge when cleaning a unit (see the call in Algorithm~\ref{alg:errorHandling} line~\ref{cleanUnit}).
	
\begin{figure}[ht]
		\begin{center}
			\begin{subfigure}[b]{0.4\textwidth}
				\includegraphics[width=\textwidth]{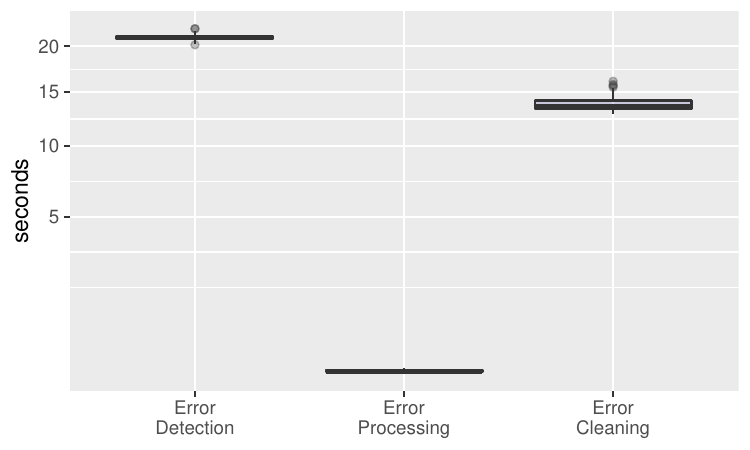}
				\caption{Virtual Machines Running on Azure}
				\label{subfig:error_azure}
			\end{subfigure}
			\\
			\begin{subfigure}[b]{0.4\textwidth}
				\centering
				\includegraphics[width=\textwidth]{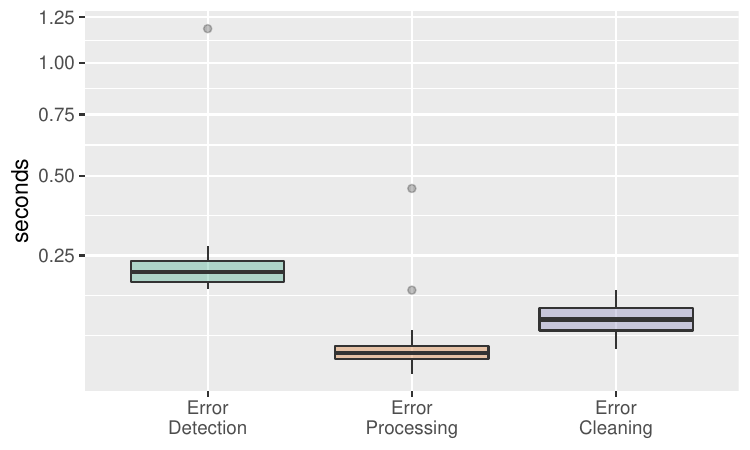}
				\caption{Containers Running on Kubernetes}
				\label{subfig:error_kubernetes}
			\end{subfigure}
			\caption{Error Handling}
			\label{fig:error_handling}
		\end{center}
		\vspace{-1.5em}
	\end{figure}

	We repeated measurements 10 times to collect stable time figures. Figure~\ref{fig:error_handling} shows the collected time figures with a semilogarithmic scale considering both virtual machines (Figure~\ref{subfig:error_azure}) and containers (Figure~\ref{subfig:error_kubernetes}).
	
	In both environments, error detection and error cleaning are more expensive than error processing. In fact, error detection requires performing the deployment, at least partially, and similarly error cleaning requires disposing monitoring units and creating new stable units, when possible. 
	
	
	Similarly to probe deployment, error handling is significantly more efficient with containers than virtual machines. For instance, error detection requires around 21 seconds with virtual machines while it can be completed in less than 0.25 seconds with containers. Similarly, error cleaning requires around 13 seconds with virtual machines, while it can be completed in about 0.15 seconds with containers, but it is important to remark that the cleaning phase with containers does not require recreating a monitoring unit that is instead only disposed. The entire error handling process can be completed in around 35 seconds with virtual machines and less than a second with containers. 
	
	\smallskip
	\textbf{Answer to RQ2} Results show how the proposed MaaS solution that flexibly allocates and destroys resources, although usable with both virtual machines and containers, are naturally more suitable for containers where errors can be recovered in seconds. 
	

	\subsection{RQ3 - How does the \deleted{architecture}\added{framework} scale for an increasing number of requests?}
	
	As discussed, the \deleted{architecture}\added{framework} can update multiple monitoring units in parallel as long as a sufficient number of controller instances are created. We thus focus the scalability study on measuring how the cost of collecting additional indicators grows with an increasing number of requests when single instances of the controllers are available. In particular, we consider two cases: processing requests that require \emph{deploying new probes} and processing requests that require \emph{reconfiguring} the monitoring system without deploying new probes. The former case corresponds to operators asking for new indicators to be collected. The latter case corresponds to operators asking for indicators already collected by other operators that the \deleted{architecture}\added{framework} handles in an optimized way sharing the existing probes among operators without touching the monitoring units, but only changing the set of configurations associated with a unit. 
	
	We measure how the total deployment time grows while increasing either the number of \emph{new indicators} or the number of \emph{existing indicators for new operators} from 1 to 30. We submit all requests at once and we measure the total time necessary to fulfill the request. We repeated every experiment 5 times on both virtual machines and containers for a total of 160 samples collected about scalability.
	
\begin{figure}[ht]
		\begin{center}
				\includegraphics[width=0.4\textwidth]{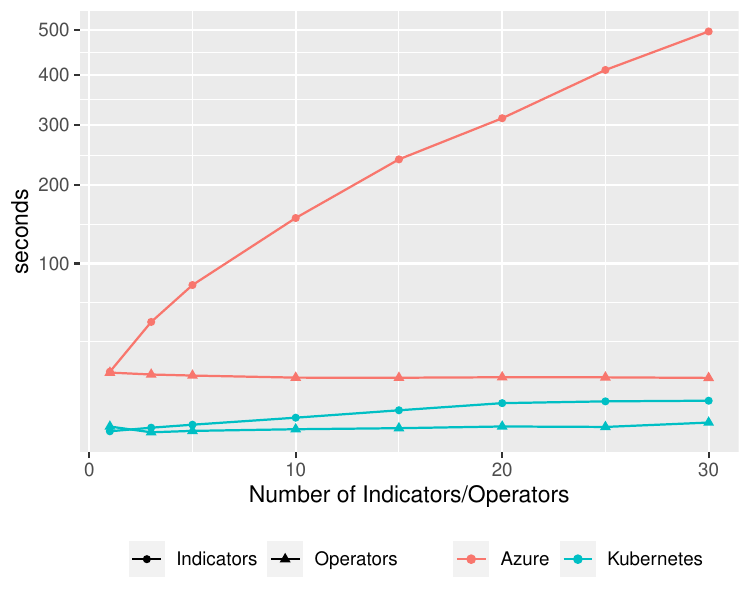}
				\caption{Scalability results.}
				\label{fig:scalability_indicators}
		\end{center}
		\vspace{-1.5em}
	\end{figure}

	Figure~\ref{fig:scalability_indicators} shows the results. Again, the remarkable difference between virtual machines and containers is confirmed. The scalability experiment gives additional evidence of how the linear growth of the total time for virtual machines is far more steep than containers. The difference is dramatic when considering the deployment of 30 probes, which requires around 10 minutes, in contrast with containers that can complete this operations in seconds.

	The results show that sharing probes between multiple operators can significantly improve the efficiency of the monitoring system. This is particularly important for virtual machines where the probe deployment cost can be cut thanks to probes sharing.

	\smallskip
	\textbf{Answer to RQ3}
	Overall, results show that dynamic probe deployment can be feasible with both virtual machines and containers. However, the former environment can efficiently deal with probes only if changes are sporadic and the number of parallel requests received is limited. On the contrary, the container technology is definitely ready to support dynamic probe deployment, even in rapidly evolving contexts, based on the proposed \deleted{architecture}\added{framework}. 
	
	\subsection{Threats to Validity}
	The threats to the validity of the results mainly concern with the relationship between the technical setup of the experiment and the collected time figures. In fact, efficiency is affected by both the available computational resources and the choice of the probes. However, while changing the available computational resources and the deployed probes are likely to affect absolute figures, the trends and gaps between the different \deleted{architecture}\added{framework}s and cloud platforms are clear, despite these factors. \changetxt{R3.C2}{In fact, plots for virtual machines and containers are similar, although values are on different scales. Further, the scalability trends clearly identify a single case (collecting increasingly more indicators on virtual machines) that scales remarkably worst than the others.}
	
	In our evaluation, we also selected a specific target service to be monitored (i.e., PostgreSQL) and we also used a specific malfunctioning probe (Metricbeat for PostgreSQL). Both these choices do not likely affect our results. In fact, the cost of handling a monitoring unit does not depend on the monitoring target, and similarly the error handling policy is the same for every type of error and malfunctioning probe.  
	
	Finally, the collected time figures might be affected by noise. To mitigate this issue we repeated experiments between 5 and 10 times. \changetxt{R3.C2}{The reported boxplots show a low variance for the collected values, suggesting that measures are stable and meaningful, and can be used to derive valid conclusions}.
	
	\section{Related Work}\label{sec:related_work}
	

	Due to the size, complexity, and dynamicity of cloud systems, Monitoring-as-a-Service (MaaS) systems are increasingly studied to better cope with the requirements of the cloud~\cite{Meng:TC:2013,duan2015xaassurvey}. In this paper, we focused on the inability of the monitoring solutions to cope with the dynamic deployment of probe artifacts, which is required to effectively address scenarios that imply quickly or frequently changing the collected indicators.
	
	
	Some MaaS systems are designed to operate \emph{tightly coupled with the target cloud technology}. Although the knowledge of the target platform may simplify the design of the MaaS solution, the resulting architecture lacks generality and is inherently limited to the collection of platform-level indicators, missing to cope with application-level indicators. For instance, MonPaaS~\cite{calero2014monpaas} is an open-source monitoring solution \added{tightly coupled with OpenStack that provides a MaaS model to cloud users. It collects indicators by integrating Nagios with OpenStack and it obtains VM status information by intercepting messages in the OpenStack message queue. Moreover, several commercial cloud monitoring solutions are developed to monitor resources running on specific cloud platforms, such as Amazon CloudWatch~\cite{amazon_2021_cloudwatch} and Google Cloud Monitoring~\cite{googlecloudmonitoring}. As a result, these solutions have limited interoperability and applicability. On the other hand, general purpose monitoring tools like Prometheus~\cite{prometheus} and Zabbix~\cite{zabbix} struggle with adapting to changing needs.
		In contrast, the framework proposed in this paper provides automation while relying on a technology-agnostic architecture that can operate probes of different type.}
	
	On the other hand, \added{MaaS systems are normally limited to \emph{interactions with pre-deployed probes}, which can be activated and deactivated by operators, but cannot be deployed/undeployed. 
		For instance, CLAMS is a MaaS framework that can monitor, benchmark and profile applications hosted on multi-clouds environments~\cite{Alhamazani:CLAMS:2014,Alhamazani:CLAMBS:2019}. 
		MEASURE is a Monitoring-as-a-service (MaaS) framework to monitor the cloud using stream processing~\cite{Smit:MISURE:2013} that relies on a publish-subscribe architecture to push data from resources through stream processors that convert and deliver data to stream subscribers. \changetxtNoRev{AdaptiveMon~\cite{Colombo:AdaptiveMon:SEAMS:2022} is a peer-to-peer monitoring solution that exploits reconfigurable probes to address the complexity of the fog environment.} Unlike our framework, these approaches rely on pre-deployed probes.}	
	
	In some cases, MaaS systems use probes that can address \emph{autoscalable services}. For instance, Amazon CloudWatch~\cite{amazon_2021_cloudwatch} uses the MaaS technology to deliver status monitoring capabilities in presence of autoscalable services. 
	JCatascopia~\cite{trihinas2014jcatascopia} is a monitoring solution that targets elastic tasks using automatic discovery capabilities. 
	Compared to these solutions, in addition to managing probes that have native support to services autoscaling and discovery, our \deleted{architecture}\added{framework} delivers automatic and declarative probe deployment capabilities, allowing users to continuously and efficiently adapt the monitoring infrastructure.
	
	\smallskip
	Early contributions related to automatic probe deployment for cloud systems are the work by Ciuffoletti~\cite{Ciuffoletti2016}, who proposed a MaaS model for the deployment of monitoring components as required by users,  
	and Anisetti et al.~\cite{Anisetti:CSC:2017}, who applied automatic probe deployment to monitor and certify security properties of services running on virtual machines. These works describe high-level design and prototype implementation targeting specific cases. 
	The \added{framework} presented in this paper provides instead a general and applicable approach for dynamic probes deployment. 

	\section{Conclusions}\label{sec:conclusions}
	
	Monitoring systems must be able to cope with dynamically changing and unstable sets of monitored indicators, to effectively address scenarios that characterize modern cloud systems. Current solutions however badly adapt to these scenarios, providing little flexibility and requiring significant manual effort to deploy, undeploy and re-configure monitoring probes.
	
	In this paper, we present a framework that can be used to dynamically work with monitoring units and probes. The operator interacts with the monitoring system in a as-a-service fashion, specifying the indicators that must be collected, and letting the framework to deal with probe deployment and configuration. The framework is also able to recover from deployment problems and integrate probes from multiple monitoring technologies.
	
	Results show that the framework can be feasibly used with cloud systems based on both virtual machines and containers, although it is significantly more efficient with containers. 
	
	\changetxt{R1.C1, R1.C2, R2.C3}{We identify three main limitations of the current framework implementation that we want to address as part of future work. First, fine-grained control of the probes configurations (e.g., changing the sampling rate of the individual probes) is not supported. This limitation can be potentially addressed by enriching monitoring claims with information about probe configurations. Second, the support to elasticity right now depends on the probe intelligence (e.g., it requires the probes to embed a discovery mechanism as the one in the Metricbeat Kubernetes module). It would be interesting to move this support at the level of the monitoring framework, so that any probe can be used to monitor elastic services. Third, error-handling is limited to the deployment phase, and it is unable to detect and repair run-time errors that occur during the regular execution of the monitoring system. Indeed, error handling requires further research to cover the full range of situations.
		
	Other future work involves extending the framework with as-a-service anomaly detection and healing capabilities to increase the dependability of the monitored system, and caring of the vicinity of the monitored resources to further improve the deployment of the monitoring system.}
	
	
	\bibliographystyle{IEEEtran}
	\bibliography{references}

\begin{thebibliography}{10}
\providecommand{\url}[1]{#1}
\csname url@samestyle\endcsname
\providecommand{\newblock}{\relax}
\providecommand{\bibinfo}[2]{#2}
\providecommand{\BIBentrySTDinterwordspacing}{\spaceskip=0pt\relax}
\providecommand{\BIBentryALTinterwordstretchfactor}{4}
\providecommand{\BIBentryALTinterwordspacing}{\spaceskip=\fontdimen2\font plus
\BIBentryALTinterwordstretchfactor\fontdimen3\font minus
  \fontdimen4\font\relax}
\providecommand{\BIBforeignlanguage}[2]{{%
\expandafter\ifx\csname l@#1\endcsname\relax
\typeout{** WARNING: IEEEtran.bst: No hyphenation pattern has been}%
\typeout{** loaded for the language `#1'. Using the pattern for}%
\typeout{** the default language instead.}%
\else
\language=\csname l@#1\endcsname
\fi
#2}}
\providecommand{\BIBdecl}{\relax}
\BIBdecl

\bibitem{CloudCompStats2020}
{hostingtribunal.com}. (2020) 25 must-know cloud computing statistics in 2020.
  \url{https://hostingtribunal.com/blog/cloud-computing-statistics/}. [Online;
  accessed 28-April-2021].

\bibitem{Ferrer:DecentralizedCloud:CSUR:2019}
A.~J. Ferrer, J.~M. Marqu\`{e}s, and J.~Jorba, ``Towards the decentralised
  cloud: Survey on approaches and challenges for mobile, ad hoc, and edge
  computing,'' \emph{ACM Computing Surveys}, vol.~51, no.~6, 2019.

\bibitem{romano_novel_2011}
L.~Romano, D.~D. Mari, Z.~Jerzak, and C.~Fetzer, ``A {Novel} {Approach} to
  {QoS} {Monitoring} in the {Cloud},'' in \emph{2011 {First} {International}
  {Conference} on {Data} {Compression}, {Communications} and {Processing}},
  Jun. 2011, pp. 45--51.

\bibitem{cedillo2015monitoringcloud}
P.~Cedillo, J.~Jimenez-Gomez, S.~Abrahao, and E.~Insfran, ``{Towards a
  Monitoring Middleware for Cloud Services},'' in \emph{2015 IEEE International
  Conference on Services Computing}, 2015, pp. 451--458.

\bibitem{shatnawi2018chmm}
A.~Shatnawi, M.~Orr\'{u}, M.~Mobilio, O.~Riganelli, and L.~Mariani,
  ``{CloudHealth: {A} Model-Driven Approach to Watch the Health of Cloud
  Services},'' in \emph{Proceedings of the 1st International Workshop on
  Software Health (SoHeal 2018)}.\hskip 1em plus 0.5em minus 0.4em\relax
  ACM/IEEE, 2018, pp. 40--47.

\bibitem{wang2011monitoringdc}
C.~Wang, K.~Schwan, V.~Talwar, G.~Eisenhauer, L.~Hu, and M.~Wolf, ``{A Flexible
  Architecture Integrating Monitoring and Analytics for Managing Large-scale
  Data Centers},'' in \emph{Proceedings of the 8th ACM International Conference
  on Autonomic Computing}, ser. ICAC '11.\hskip 1em plus 0.5em minus
  0.4em\relax ACM, 2011, pp. 141--150.

\bibitem{kutare2010monalytics}
M.~Kutare, K.~Schwan, G.~Eisenhauer, V.~Talwar, C.~Wang, and M.~Wolf,
  ``{Monalytics: online monitoring and analytics for managing large scale data
  centers},'' in \emph{In ICAC ’10: Proceeding of the 7th international
  conference on Autonomic computing}.\hskip 1em plus 0.5em minus 0.4em\relax
  ACM, 2010, pp. 141--150.

\bibitem{kubernetes}
\BIBentryALTinterwordspacing
{The Linux Foundation}. (2021) Kubernetes. [Online; accessed 26-April-2022].
  [Online]. Available: \url{https://kubernetes.io/}
\BIBentrySTDinterwordspacing

\bibitem{elasticsearch_2020_stack}
{Elasticsearch BV}. (2021) {The Elastic Stack}. \url{
  https://www.elastic.co/elastic-stack }. [Online; accessed 26-April-2022].

\bibitem{prometheus}
{The Linux Foundation}. (2021) Prometheus. \url{https://prometheus.io/}.
  [Online; accessed 26-April-2022].

\bibitem{redhat_2020_ansible}
{Red Hat, Inc.} (2021) {How Ansible Works}. \url{
  https://www.ansible.com/overview/how-ansible-works }. [Online; accessed
  26-April-2022].

\bibitem{puppet}
{Puppet}. (2021) {Puppet}. \url{https://puppet.com }. [Online; accessed
  26-April-2022].

\bibitem{trihinas2014jcatascopia}
D.~Trihinas, G.~Pallis, and M.~D. Dikaiakos, ``Jcatascopia: Monitoring
  elastically adaptive applications in the cloud,'' in \emph{2014 14th IEEE/ACM
  International Symposium on Cluster, Cloud and Grid Computing}.\hskip 1em plus
  0.5em minus 0.4em\relax IEEE, 2014, pp. 226--235.

\bibitem{hp_2020_monasca}
{Hewlett-Packard Enterprise Development LP}. (2021) {Monasca - an {OpenStack}
  {Community} project}. \url{https://monasca.io/ }. [Online; accessed
  26-April-2022].

\bibitem{FATEMA20142918}
\BIBentryALTinterwordspacing
K.~Fatema, V.~C. Emeakaroha, P.~D. Healy, J.~P. Morrison, and T.~Lynn, ``A
  survey of cloud monitoring tools: Taxonomy, capabilities and objectives,''
  \emph{Journal of Parallel and Distributed Computing}, vol.~74, no.~10, pp.
  2918 -- 2933, 2014. [Online]. Available:
  \url{http://www.sciencedirect.com/science/article/pii/S0743731514001099}
\BIBentrySTDinterwordspacing

\bibitem{aceto2013monitoringsurvey}
G.~Aceto, A.~Botta, W.~de~Donato, and A.~Pescapè, ``{Cloud monitoring: A
  survey},'' \emph{Computer Networks}, vol.~57, no.~9, pp. 2093--2115, 2013.

\bibitem{Tundo:Varys:ESECFSE:2019}
A.~Tundo, M.~Mobilio, M.~Orr\`{u}, O.~Riganelli, M.~Guzm\`{a}n, and L.~Mariani,
  ``Varys: An agnostic model-driven monitoring-as-a-service framework for the
  cloud,'' in \emph{Proceedings of the 27th ACM Joint Meeting on European
  Software Engineering Conference and Symposium on the Foundations of Software
  Engineering (ESEC/FSE), tool demo}, 2019.

\bibitem{stallings1998snmp}
W.~Stallings, \emph{SNMP, SNMPv2, SNMPv3, and RMON 1 and 2}.\hskip 1em plus
  0.5em minus 0.4em\relax Addison-Wesley Longman Publishing Co., Inc., 1998.

\bibitem{burns2016design}
B.~Burns and D.~Oppenheimer, ``{Design patterns for container-based distributed
  systems},'' in \emph{Proceedings of the 8th USENIX Conference on Hot Topics
  in Cloud Computing}.\hskip 1em plus 0.5em minus 0.4em\relax USENIX
  Association, 2016, pp. 108--113.

\bibitem{apache_2020_kafka}
{The Apache Software Foundation}. (2021) {Apache Kafka}. \url{
  https://kafka.apache.org/ }. [Online; accessed 26-April-2022].

\bibitem{mongodb_2020_mongodb}
{MongoDB, Inc.} (2021) {MongoDB}. \url{ https://www.mongodb.com/ }. [Online;
  accessed 26-April-2022].

\bibitem{redis.io_2020_redis}
{Salvatore Sanfilippo and contributors}. (2021) {Redis.io}.
  \url{https://redis.io/}. [Online; accessed 26-April-2022].

\bibitem{elasticsearch_2020_beats}
{Elasticsearch BV}. (2021) {Beats: Data Shippers for Elasticsearch}.
  \url{https://www.elastic.co/beats/}. [Online; accessed 26-April-2022].

\bibitem{elastic_2020_elasticsearch}
------. (2021) {Elasticsearch: RESTful, Distributed Search \& Analytics}. \url{
  https://www.elastic.co/elasticsearch/ }. [Online; accessed 26-April-2022].

\bibitem{Meng:TC:2013}
S.~{Meng} and L.~{Liu}, ``Enhanced monitoring-as-a-service for effective cloud
  management,'' \emph{IEEE Transactions on Computers}, vol.~62, no.~9, pp.
  1705--1720, 2013.

\bibitem{duan2015xaassurvey}
Y.~Duan, G.~Fu, N.~Zhou, X.~Sun, N.~C. Narendra, and B.~Hu, ``{Everything as a
  Service (XaaS) on the Cloud: Origins, Current and Future Trends},'' in
  \emph{2015 IEEE 8th International Conference on Cloud Computing}, 2015, pp.
  621--628.

\bibitem{calero2014monpaas}
J.~M.~A. Calero and J.~G. Aguado, ``Monpaas: an adaptive monitoring platformas
  a service for cloud computing infrastructures and services,'' \emph{IEEE
  Transactions on Services Computing}, vol.~8, no.~1, pp. 65--78, 2014.

\bibitem{amazon_2021_cloudwatch}
{Amazon Web Services, Inc.} (2021) {CloudWatch}.
  \url{https://aws.amazon.com/cloudwatch/}. [Online; accessed 26-April-2022].

\bibitem{googlecloudmonitoring}
{Google}. (2021) Cloud monitoring | google cloud.
  \url{https://cloud.google.com/monitoring}. [Online; accessed 26-April-2022].

\bibitem{zabbix}
{Zabbix LLC}. (2021) Zabbix features overview.
  \url{https://www.zabbix.com/features}. [Online; accessed 26-April-2022].

\bibitem{Alhamazani:CLAMS:2014}
K.~{Alhamazani}, R.~{Ranjan}, K.~{Mitra}, P.~P. {Jayaraman}, Z.~{Huang},
  L.~{Wang}, and F.~{Rabhi}, ``Clams: Cross-layer multi-cloud application
  monitoring-as-a-service framework,'' in \emph{2014 IEEE International
  Conference on Services Computing}, 2014, pp. 283--290.

\bibitem{Alhamazani:CLAMBS:2019}
K.~{Alhamazani}, R.~{Ranjan}, P.~{Prakash Jayaraman}, K.~{Mitra}, C.~{Liu},
  F.~{Rabhi}, D.~{Georgakopoulos}, and L.~{Wang}, ``Cross-layer multi-cloud
  real-time application qos monitoring and benchmarking as-a-service
  framework,'' \emph{IEEE Transactions on Cloud Computing}, vol.~7, no.~1, pp.
  48--61, 2019.

\bibitem{Smit:MISURE:2013}
M.~Smit, B.~Simmons, and M.~Litoiu, ``Distributed, application-level monitoring
  for heterogeneous clouds using stream processing,'' \emph{Future Gener.
  Comput. Syst.}, vol.~29, no.~8, 2013.

\bibitem{Colombo:AdaptiveMon:SEAMS:2022}
V.~Colombo, A.~Tundo, M.~Ciavotta, and L.~Mariani, ``{Towards Self-Adaptive
  Peer-to-Peer Monitoring for Fog Environments},'' in \emph{Proceedings of the
  17th Symposium on Software Engineering for Adaptive and Self-Managing Systems
  (SEAMS)}, 2022.

\bibitem{Ciuffoletti2016}
A.~Ciuffoletti, ``Application level interface for a cloud monitoring service,''
  \emph{Computer Standards \& Interfaces}, vol.~46, pp. 15 -- 22, 2016.

\bibitem{Anisetti:CSC:2017}
M.~{Anisetti}, C.~A. {Ardagna}, E.~{Damiani}, and F.~{Gaudenzi}, ``A
  semi-automatic and trustworthy scheme for continuous cloud service
  certification,'' \emph{IEEE Transactions on Services Computing}, vol.~13,
  no.~1, pp. 30--43, 2017.

\end{thebibliography}
	
	\begin{IEEEbiography}[{\includegraphics[width=1in,height=1.25in,clip,keepaspectratio]{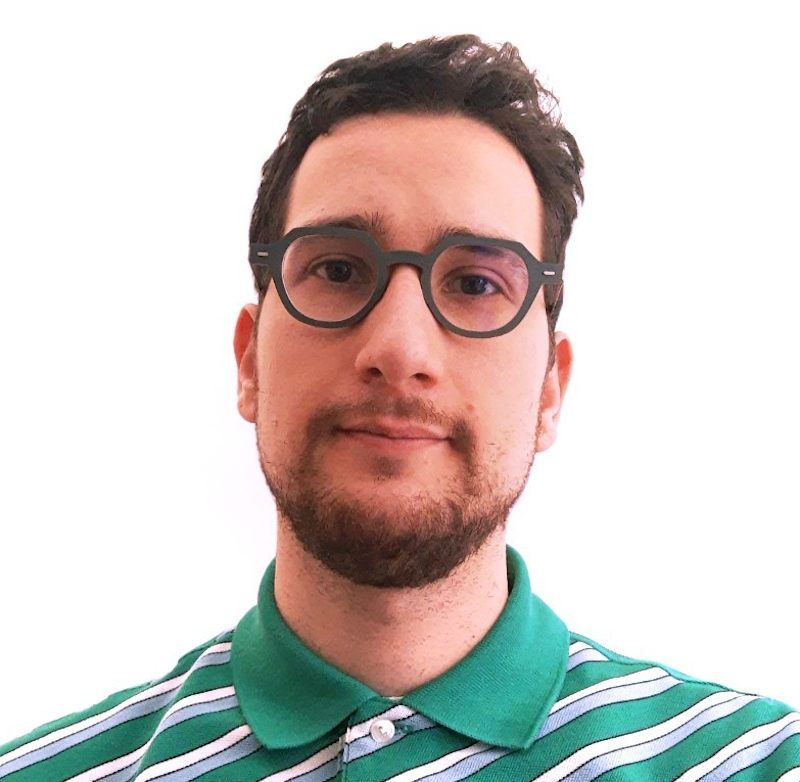}}]{Alessandro Tundo} is a Ph.D. student at the University of Milano-Bicocca. He holds a Master Degree in Computer Science received from the same university in 2018.
		
		His research interests include cloud and fog computing, distributed systems monitoring and software architectures. \end{IEEEbiography}
	
	\begin{IEEEbiography}[{\includegraphics[width=1in,height=1.25in,clip,keepaspectratio]{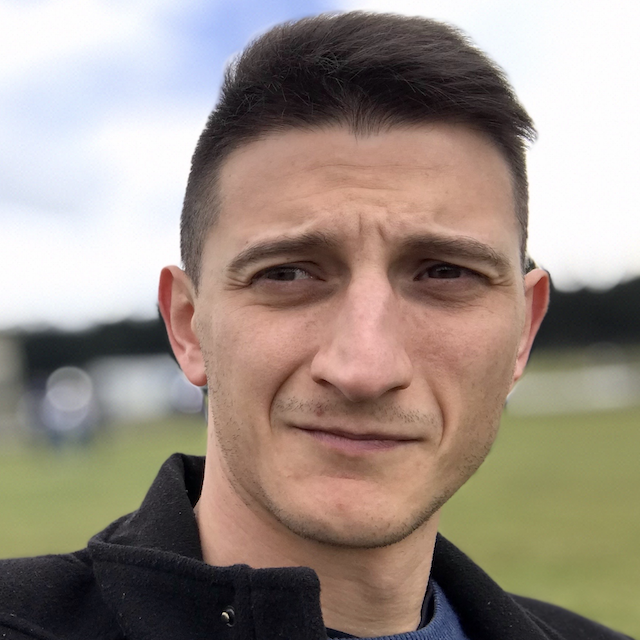}}]{Marco Mobilio} is a post-doc at the University of Milano-Bicocca, where he got his Ph.D. in 2017 and his Master Degree in Computer Science in 2013. His main interests cover Software Architecture, Cloud Monitoring and Self-Healing, Automatic Testing for web and mobile applications and Human Activity Recognition. \end{IEEEbiography}
	
	\begin{IEEEbiography}[{\includegraphics[width=1in,height=1.25in,clip,keepaspectratio]{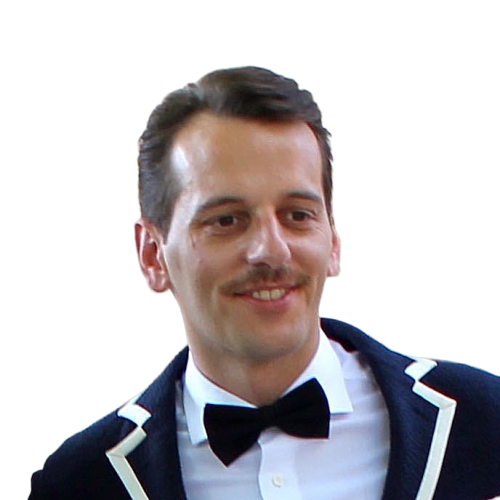}}]{Oliviero Riganelli} is an Assistant Professor at the University of Milano-Bicocca. He holds a Ph.D., in Computer Science and Complex System from the University of Camerino in 2009. He is a computer scientist with a keen interest in Software Engineering. His main research interests focus on creating advanced methodologies and technologies to build better software by automatically testing, analyzing, and correcting the software itself and its development process. He is and has been involved in several research projects, both international and national, in close collaboration with leading partners from industry and academia. He is also regularly involved in the program committees of workshops and conferences in his areas of interest\end{IEEEbiography}
	
	\begin{IEEEbiography}[{\includegraphics[width=1in,height=1.25in,clip,keepaspectratio]{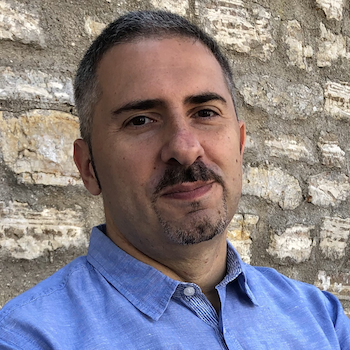}}]{Leonardo Mariani} is a Full Professor at the University of Milano-Bicocca. He holds a Ph.D. in Computer Science received from the same university in 2005.
		
		His research interests include software engineering, in particular software testing, program analysis, automated debugging, specification mining, and self-healing and self-repairing systems. He has authored more than 100 papers appeared at top software engineering conferences and journals.
		
		He has been awarded with the ERC Consolidator Grant in 2015, an ERC Proof of Concept grant in 2018, and he is currently active in several European and National projects. He is regularly involved in organizing and program committees of major software engineering conferences.
	\end{IEEEbiography}
	
\end{document}